\documentclass[12pt]{article}

\usepackage{amssymb}

\makeatletter
 \@addtoreset{equation}{section}
 
\makeatother

\begin{document}


\begin{titlepage}

\renewcommand{\thefootnote}{\fnsymbol{footnote}}

\begin{flushright}
\begin{tabular}{l}
UTHEP-613 \\
OIQP-10-10
\end{tabular}
\end{flushright}

\bigskip

\begin{center}
{\Large \bf 
Light-cone Gauge NSR Strings in Noncritical Dimensions II 
---Ramond Sector--- 
\\}
\end{center}

\bigskip

\begin{center}
{\large Nobuyuki Ishibashi}${}^{a}$\footnote{e-mail:
        ishibash@het.ph.tsukuba.ac.jp}
and
{\large Koichi Murakami}${}^{b}$\footnote{e-mail:
        koichimurakami71@gmail.com}
\end{center}

\begin{center}
${}^{a}${\it
Institute of Physics, University of Tsukuba,\\
Tsukuba, Ibaraki 305-8571, Japan}\\
\end{center}
\begin{center}
$^{b}$\textit{Okayama Institute for Quantum Physics,}\\
\textit{ Kyoyama 1-9-1, Kita-ku, Okayama 700-0015, Japan} 
\par\end{center}

\bigskip

\bigskip

\bigskip

\begin{abstract}
Light-cone gauge superstring theory in noncritical dimensions corresponds
to a worldsheet theory with nonstandard longitudinal part in the conformal
gauge. The longitudinal part of the worldsheet theory is a superconformal
field theory called $X^{\pm}$ CFT.
We show that the $X^{\pm}$ CFT combined with the super-reparametrization
ghost system can be described by free variables. It is possible to
express the correlation functions in terms of these free variables.
Bosonizing the free variables, we construct the spin fields and
BRST invariant vertex operators for the Ramond sector in the conformal
gauge formulation. 
By using these vertex operators, we can rewrite 
the tree amplitudes of the noncritical light-cone gauge string field theory,
with external lines in the (R,R) sector as well as those in the (NS,NS)
sector, in a BRST invariant way. 
\end{abstract}

\setcounter{footnote}{0}
\renewcommand{\thefootnote}{\arabic{footnote}}

\end{titlepage}

\section{Introduction}

The light-cone gauge string field theory~\cite{Mandelstam:1973jk,
Kaku:1974zz,Kaku:1974xu,Cremmer:1974ej,Mandelstam:1974hk,Sin:1988yf}
takes a simple form and it can therefore be a very useful tool to
study string theory. Being a gauge fixed theory, it can be formulated
in noncritical spacetime dimensions. In the conformal gauge, such
noncritical string theories correspond to worldsheet theories with nonstandard
longitudinal part. In our previous works~\cite{Baba:2009ns,Baba:2009fi},
we have studied the longitudinal part of the worldsheet theory which
is an interacting CFT called $X^{\pm}$ CFT. It has the right value
of the Virasoro central charge, so that one can construct a nilpotent
BRST charge combined with the transverse part and the reparametrization
ghosts.

In the conformal gauge formulation, the amplitudes can be calculated
in a BRST invariant manner. In Ref.~\cite{Baba:2009ns} we have shown
that the tree level amplitudes in the light-cone gauge coincide with
the BRST invariant ones in the conformal gauge, in the case of the
bosonic noncritical strings. For superstrings, the equivalence of
the amplitudes in the two gauges has been shown for the cases where
all the external lines are in the (NS,NS) sector%
\footnote{In our previous works and in this work, 
we discuss closed strings.}~\cite{Baba:2009zm}.

We would like to extend this analysis into the case in which external
lines in the Ramond sector are involved. In the conformal gauge formulation,
the vertex operators corresponding to the external lines in the Ramond
sector should involve the spin fields in the $X^{\pm}$ CFT. Since
the $X^{\pm}$ CFT is an interacting theory, it is not straightforward
to construct spin fields. In this paper, we formulate a free field
description of the CFT consisting of the $X^{\pm}$ CFT and the reparametrization
ghosts. Namely, we construct free variables which can be expressed
in terms of $X^{\pm}$ and the ghosts. We provide a formula to express
the correlation functions of this interacting CFT in terms of these
free variables. Bosonizing the free variables, we define the spin
fields and thereby construct the vertex operators in the Ramond sector.

In the conformal gauge, the amplitudes can be expressed by 
the vertex operators thus constructed. 
It turns out that the closed superstring theory in 
noncritical dimensions generically does not include spacetime fermions.
We show that the tree amplitudes with not only external lines in the (NS,NS) sector
but also those in the (R,R) sector  can be written by using the vertex operators.

This paper is organized as follows. In section~\ref{sec:Free-field-realization},
we consider the system of the bosonic $X^{\pm}$ CFT combined with the 
reparametrization ghosts and construct free variables. 
We show how the correlation functions on the complex plane can be
expressed by those of the free variables. 
In section \ref{sec:supersymmetric-case},
we supersymmetrize the analyses in section~\ref{sec:Free-field-realization}
and formulate the free field description of the supersymmetric $X^{\pm}$
CFT. In section~\ref{sec:vo}, we first study how the BRST invariant
vertex operators in the Neveu-Schwarz sector can be described in terms
of free variables obtained in section~\ref{sec:supersymmetric-case}.
Then we construct those in the Ramond sector, using the free variables. 
In section\ \ref{sec:Amplitudes}, we show that the
tree amplitudes involving external lines in the (R,R) and the (NS,NS)
sectors of the noncritical strings can be expressed in a BRST invariant
way using the BRST invariant vertex operators. 
Section~\ref{sec:Conclusions-and-discussions} is devoted to 
conclusions and discussions. 
In appendix~\ref{sec:SFTaction},
we explain some details of the action for the strings 
in the (R,R) and the (NS,NS) sectors of the light-cone gauge 
string field theory in noncritical dimensions. 
In appendix~\ref{sec:method2}, we present a proof of 
a relation which we use in section~\ref{sec:Amplitudes}.


\section{Free variables: bosonic case 
         \label{sec:Free-field-realization}}

As a warm-up, we would like to present the free field description
for bosonic $X^{\pm}$ CFT formulated in Ref.~\cite{Baba:2009ns}
and show how the correlation functions are expressed by using the
free variables.

\subsection{Bosonic $X^{\pm}$ CFT}

In the conformal gauge, the longitudinal part of the worldsheet theory
for the noncritical light-cone gauge string theory is described by
a conformal field theory with the energy-momentum tensor
\begin{equation}
 \partial X^{+}\partial X^{-}-\frac{d-26}{12}\left\{ X^{+},z\right\} \ ,
\label{eq:Xpmem}
\end{equation}
where 
\begin{equation}
\left\{ X^{+},z\right\} 
  \equiv
  \frac{\partial^{3}X^{+}}{\partial X^{+}}
  -\frac{3}{2}
    \left( \frac{\partial^{2}X^{+}}{\partial X^{+}} \right)^{2}
\end{equation}
is the Schwarzian derivative.

Such a conformal field theory can be studied using the path integral
formalism \cite{Baba:2009ns}. In order to make the theory well-defined,
we always consider the situations where the vertex operators of the
form $e^{-ip^{+}X^{-}}$ are inserted so that $\partial X^{+}$ has
an expectation value and it is invertible except for sporadic points
on the worldsheet. Indeed, for a functional $F[X^{+}]$ of $X^{+}$,
one can calculate the correlation function with the insertion 
$\prod_{r=1}^{N}e^{-ip_{r}^{+}X^{-}}\left(Z_{r},\bar{Z}_{r}\right)$
on the complex plane as 
\begin{equation}
\left\langle 
  F\left[X^{+}\right]
  \prod_{r=1}^{N}e^{-ip_{r}^{+}X^{-}} \left(Z_{r},\bar{Z}_{r}\right)
\right\rangle
 =F  \left[ -\frac{i}{2}\left(\rho+\bar{\rho}\right) \right]
  \left\langle 
      \prod_{r=1}^{N}e^{-ip_{r}^{+}X^{-}} \left(Z_{r},\bar{Z}_{r}\right)
  \right\rangle \ ,
\label{eq:FX+}
\end{equation}
where
\begin{equation}
\rho(z)=\sum_{r=1}^{N}\alpha_{r}\ln\left(z-Z_{r}\right)~,
 \qquad
 \alpha_{r}\equiv2p_{r}^{+}\ .
\label{eq:mandelstamN}
\end{equation}
Thus one can see that $X^{+}$ acquires an expectation value 
$-\frac{i}{2 }\left(\rho \left(z\right) 
                    +\bar{\rho}\left(\bar{z}\right)
              \right)$.
The expectation value of $\partial X^{+}\left(z\right)$ is proportional
to $\partial\rho\left(z\right)$. 
$\partial\rho\left(z\right)$ has $N$ poles at $z=Z_{r}$ 
and $N-2$ zeros at $z=z_{I}\ \left(I=1,\cdots N-2\right)$.
$\rho\left(z\right)$ coincides with the Mandelstam mapping of a tree
light-cone diagram for $N$ strings and $z_{I}$ are the interaction points.

The variables $X^{\pm}$ can be shown to satisfy the OPE's
\begin{eqnarray}
\partial X^{+}(z)\partial X^{+}(z') 
  & \sim & \mathrm{regular}\ ,
\nonumber \\
\partial X^{-}(z)\partial X^{+}(z') 
  & \sim & \frac{1}{(z-z')^{2}}~,
\nonumber \\
\partial X^{-}(z)\partial X^{-}(z') 
  & \sim & -\frac{d-26}{12}\partial_{z}\partial_{z'}
           \left[ \frac{1}{\left(X_{L}^{+}(z)-X_{L}^{+}(z')\right)^{2}}
           \right]\ ,
\label{eq:XpmOPE}
\end{eqnarray}
where $X_{L}^{+}$ denotes the left-moving part of $X^{+}$. 
Expanding the right hand side of the third equation in terms of 
$z-z^{\prime}$ with the assumption $\left|z-z^{\prime}\right|\ll1$, 
one gets the form of the OPE given in Ref.~\cite{Baba:2009ns}. 
Using these OPE's,
one can show that the energy-momentum tensor (\ref{eq:Xpmem}) satisfies
the Virasoro algebra with central charge $28-d$. Thus, with the
reparametrization ghosts and the transverse part, the worldsheet theory becomes a CFT
with the total central charge $0$.

\subsection{Free fields}

Let us consider a 2D CFT which consists of the $X^{\pm}$ CFT and
the system of reparametrization ghosts $b,c,\tilde{b},\tilde{c}$.
One can show that this theory can be described by 
free variables~\cite{Baba:2009ns}.
Free variables 
$X^{+},X^{\prime-},$ 
$b^{\prime},c^{\prime}$, 
$\tilde{b}^{\prime},\tilde{c}^{\prime}$
are defined as 
\begin{eqnarray}
 &  & \quad 
      b^{\prime} \equiv \left(\partial X^{+}\right)^{\alpha}b\ ,
      \qquad
      \,\,
      \tilde{b}^{\prime}
         \equiv \left(\bar{\partial}X^{+}\right)^{\alpha}
                \tilde{b}\ ,
\nonumber \\
 &  & \quad 
      c^{\prime} \equiv \left(\partial X^{+}\right)^{-\alpha}c\ ,
      \qquad
      \tilde{c}^{\prime}
          \equiv \left(\bar{\partial}X^{+}\right)^{-\alpha}
                 \tilde{c}\ ,
\nonumber \\
 &  & X^{\prime-} 
     \equiv X^{-}
       -\alpha\frac{cb}{\partial X^{+}}
       -\frac{3}{2} \alpha
        \frac{\partial^{2}X^{+}}{\left(\partial X^{+}\right)^{2}}
       -\alpha\frac{\tilde{c}\tilde{b}}{\bar{\partial}X^{+}}
       -\frac{3}{2}\alpha
         \frac{\bar{\partial}^{2}X^{+}}
              {\left(\bar{\partial}X^{+}\right)^{2}}\ ,
\label{eq:free}
\end{eqnarray}
with 
\begin{equation}
\alpha\left(\alpha+3\right)=\frac{d-26}{12}\ .
\label{eq:alpha-bosonic}
\end{equation}
The OPE's between 
 $X^{+},X^{\prime-},b^{\prime},c^{\prime},
   \tilde{b}^{\prime},\tilde{c}^{\prime}$
can be derived from the OPE's of $X^{\pm},b,c,\tilde{b},\tilde{c}$
and one can see that they are free variables. It is straightforward
to show that the energy-momentum tensor of the system 
\begin{equation}
 T\left(z\right) 
  =\partial X^{+}\partial X^{-}
    -\frac{d-26}{12}\left\{ X^{+},z\right\} 
    -2b\partial c-\partial bc\ ,
\label{eq:EM-bosonic}
\end{equation}
can be written as 
\begin{equation}
T\left(z\right)
 =\partial X^{+}\partial X^{\prime-}
  -b^{\prime}\partial c^{\prime}
  -\left(1+\alpha\right)\partial\left(b^{\prime}c^{\prime}\right)\ ,
\label{eq:EM-bosonic2}
\end{equation}
in the form of the energy-momentum tensor for the free fields 
$X^{+},X^{\prime-}$,
$b^{\prime},c^{\prime}$. 
The fields $b^{\prime},c^{\prime}$ are with conformal
weight $\left(2+\alpha,0\right),\left(-1-\alpha,0\right)$ respectively.
It is also easy to express $X^{\pm},b,c,\tilde{b},\tilde{c}$ in terms
of the free variables.

\subsection{Correlation functions}

Since one can express all the fields in the theory in terms of the
free variables and vice versa, it should be possible to describe the
theory using these free variables. Let 
\begin{equation}
\left\langle   \phi_{1}\phi_{2}\cdots\phi_{N}
\right\rangle _{X^{\pm},b,c}
\end{equation}
denote the correlation function on the complex plane
in the CFT we are considering. 
As we mentioned above, in the $X^{\pm}$ CFT, we are mainly interested
in the the correlation functions with insertions of 
$e^{-ip^{+}X^{-}}$.
In our setup, the correlation functions to be considered are 
of the form 
\begin{equation}
\left\langle 
  \left| e^{3\sigma }\left(\infty\right) \right|^{2}
   \phi_{1}\left(z_{1},\bar{z}_{1}\right)
   \phi_{2}\left(z_{2},\bar{z}_{2}\right)
    \cdots
   \phi_{n}\left(z_{n},\bar{z}_{n}\right)
   \prod_{r=1}^{N}e^{-ip_{r}^{+}X^{-}}\left(Z_{r},\bar{Z}_{r}\right)
\right\rangle _{X^{\pm},b,c}\ ,
\label{eq:phi1cdotsphin}
\end{equation}
where $\partial\sigma=cb$ 
and $\phi_{i}\ \left(i=1,\cdots,n\right)$
are local operators made from 
$X^{+},\partial X^{-},\bar{\partial}X^{-},b,c,\tilde{b},\tilde{c}$
and their derivatives. 
$\left|e^{3\sigma}\left(\infty\right)\right|^{2}$
is inserted to soak up the ghost zero modes.

The correlation function (\ref{eq:phi1cdotsphin}) should be expressed
by using the free variables. Let us define the correlation function
for the free theory on the complex plane as
\begin{equation}
\left\langle 
    \phi_{1}\phi_{2}\cdots\phi_{n}
\right\rangle _{\mathrm{free}}
   \equiv
     \frac{\int \left[dX^{+}dX^{\prime-}db^{\prime}dc^{\prime}
                      d\tilde{b}^{\prime}d\tilde{c}^{\prime}\right]
                 e^{-S_{\mathrm{free}}
                     \left[X^{+},X^{\prime-},b^{\prime},c^{\prime},
                          \tilde{b}^{\prime},\tilde{c}^{\prime}\right]}
                \phi_{1} \phi_{2} \cdots \phi_{n}}
          {\int \left[ dX^{+}dX^{\prime-}db^{\prime}dc^{\prime}
                       d\tilde{b}^{\prime}d\tilde{c}^{\prime} \right]
                e^{-S_{\mathrm{free}}
                    \left[ X^{+},X^{\prime-},b^{\prime},c^{\prime},
                          \tilde{b}^{\prime},\tilde{c}^{\prime} 
                    \right]}}\ .
\end{equation}
Naively, one might expect that the correlation
function (\ref{eq:phi1cdotsphin}) should be expressed in terms of
the free variables as 
\begin{eqnarray}
\lefteqn{
  \left\langle 
        \left|e^{3\sigma}\left(\infty\right)\right|^{2}
        \phi_{1}\left(z_{1},\bar{z}_{1}\right)
        \phi_{2}\left(z_{2},\bar{z}_{2}\right)
        \cdots
        \phi_{n}\left(z_{n},\bar{z}_{n}\right)
        \prod_{r=1}^{N}
            e^{-ip_{r}^{+}X^{-}}\left(Z_{r},\bar{Z}_{r}\right)
  \right\rangle _{X^{\pm},b,c}
}\nonumber \\
 &  & =\left\langle 
         \left| e^{3\sigma}\left(\infty\right) \right|^{2}
         \phi_{1}\left(z_{1},\bar{z}_{1}\right)
         \phi_{2}\left(z_{2},\bar{z}_{2}\right)
         \cdots
         \phi_{n}\left(z_{n},\bar{z}_{n}\right)
         \prod_{r=1}^{N}
            e^{-ip_{r}^{+}X^{-}}\left(Z_{r},\bar{Z}_{r}\right)
     \right\rangle _{\mathrm{free}}\ ,
\label{eq:naive}
\end{eqnarray}
on the right hand side of which $\sigma$, $\phi_{i}$ and $X^{-}$
are considered to be expressed by the free variables 
using the relations~(\ref{eq:free}). 
Eq.(\ref{eq:naive}) would hold if the relations (\ref{eq:free}) 
were not singular anywhere on the complex plane. 
However,
if the expectation value of $\partial X^{+}$ has zeros and poles,
the relations (\ref{eq:free}) are not well-defined at these points.
Then we need to modify eq.(\ref{eq:naive}) and insert operators at
these points on the right hand side.

\subsection{Operator insertions}

The necessity of such insertions can be seen by considering the case
where all the $\phi_{i}$ do not involve derivatives of $X^{-}$ in
eq.(\ref{eq:naive}). Because of eq.(\ref{eq:FX+}), $X^{+}$ in the
correlation function can be replaced by its expectation value 
$-\frac{i}{2}\left(\rho+\bar{\rho}\right)$
in such a case. 

If $\partial X^{+}$ is replaced by $-\frac{i}{2}\partial\rho$, 
the relations between the ghost variables are the ones 
which were studied in 
Refs.~\cite{D'Hoker:1987pr,Kunitomo:1987uf,D'Hoker:1989ae}. 
They showed that the correlation functions of $b,c$ 
can be expressed by those of $b^{\prime},c^{\prime}$ 
with extra operator insertions at $z=z_{I},Z_{r},\infty$. 
For example, if $b\left(z\right)$, $c\left(z\right)$
are regular at $z=z_{I}$,%
\footnote{Here we assume in eq.(\ref{eq:naive}) 
  the generic configuration 
  in  which $z_{i}\ne Z_{r},z_{I}\ \left(i=1,\cdots,n\right)$. 
  The special cases where $z_{i}$ coincides with one of these points
  are realized as a limit of the generic ones.}
the relation (\ref{eq:free}) implies that 
$b^{\prime}\left(z\right)$, $c^{\prime}\left(z\right)$
are singular at $z=z_{I}$, because $\partial\rho\left(z_{I}\right)=0$.
One can see 
\begin{equation}
b^{\prime}\left(z\right) \sim \left(z-z_{I}\right)^{\alpha}\ ,
\qquad 
c^{\prime}\left(z\right) \sim \left(z-z_{I}\right)^{-\alpha}\ ,
\label{eq:bprimecprime}
\end{equation}
for $z\sim z_{I}$. 
Such singularities are induced by the insertions
$e^{-\alpha\sigma^{\prime}}\left(z_{I}\right)$, 
where $\sigma^{\prime}\left(z\right)$
is defined so that $\partial\sigma^{\prime}=c^{\prime}b^{\prime}$.
Therefore the correlation functions of $b, c$ with no insertions at
$z=z_{I}$ should correspond to those of $b^{\prime}, c^{\prime}$
with insertions of $e^{-\alpha\sigma^{\prime}}\left(z_{I}\right)$.
Thus we can see that eq.(\ref{eq:naive}) cannot be true as it is.
It should at least be modified as 
\begin{eqnarray}
\lefteqn{
  \left\langle 
     \left| e^{3\sigma}\left(\infty\right) \right|^{2}
     \phi_{1} \phi_{2} \cdots \phi_{n}
     \prod_{r=1}^{N}
         e^{-ip_{r}^{+}X^{-}}\left(Z_{r},\bar{Z}_{r}\right)
  \right\rangle _{X^{\pm},b,c}
}\nonumber \\
& & \sim
     \left\langle 
       \left| e^{3\sigma}\left(\infty\right) \right|^{2}
       \phi_{1} \phi_{2} \cdots \phi_{n}
       \prod_{I} \left| e^{-\alpha\sigma^{\prime}} \left(z_{I}\right)
                 \right|^{2}
       \prod_{r=1}^{N}
           e^{-ip_{r}^{+}X^{-}}\left(Z_{r},\bar{Z}_{r}\right)
\right\rangle _{\mathrm{free}}\ ,\label{eq:ealphasigma}
\end{eqnarray}
in order to be consistent with the singularities 
of the ghost variables.

In our case, $X^{+}$ is dynamical and eq.(\ref{eq:ealphasigma})
is still inconsistent. 
If one inserts the energy-momentum tensor $T\left(z\right)$
into the correlation functions in eq.(\ref{eq:ealphasigma}), the
left hand side should be regular at $z=z_{I}$~\cite{Baba:2009ns}
but the right hand side is not because of 
$e^{-\alpha\sigma^{\prime}}\left(z_{I}\right)$.
Instead of $e^{-\alpha\sigma'}(z_{I})$, we therefore need to insert
an operator which is conformal invariant and induces the same singularities
for $b^{\prime},c^{\prime}$ as $e^{-\alpha\sigma^{\prime}}\left(z_{I}\right)$.
We find that 
\begin{equation}
\mathcal{O}_{I}
  \equiv
  \left| \oint_{z_{I}} \frac{dz}{2\pi i }\partial\Phi
         \frac{e^{-\alpha\sigma^{\prime}}}
              {\left(\partial^{2}X^{+}
               \right)^{\frac{3}{4}\alpha\left(\alpha+1\right)}}
                 \left(z\right)
  \right|^{2}\ ,
\end{equation}
where 
\begin{equation}
\Phi \equiv
     \ln\partial X^{+}\bar{\partial}X^{+}~,
\label{eq:bosonicPhi}
\end{equation}
has such properties. 
Indeed, replacing $X^{+}$ by its expectation
value $-\frac{i}{2}(\rho+\bar{\rho})$, one can see that $\mathcal{O}_{I}$
is equivalent to 
\begin{equation}
\left| \frac{e^{-\alpha\sigma^{\prime}}}
            {\left(\partial^{2}\rho
             \right)^{\frac{3}{4}\alpha\left(\alpha+1\right)}}
               \left(z_{I}\right)
\right|^{2}\ ,
\end{equation}
and $b^{\prime},c^{\prime}$ behave as eq.(\ref{eq:bprimecprime})
in the presence of $\mathcal{O}_{I}$. 
Moreover, the OPE with the energy-momentum tensor can
be calculated as 
\begin{eqnarray}
T\left(z\right)\mathcal{O}_{I}
  & \sim & 
    \oint_{z_{I} }\frac{dw}{2\pi i}\,
      \partial_{w} \left[
           \frac{1}{z-w}
           \left( \partial\left(\ln\partial X^{+}\right)
                 \frac{e^{-\alpha\sigma^{\prime}}}
                      {\left(\partial^{2}X^{+}
                       \right)^{\frac{3}{4}\alpha\left(\alpha+1\right)}}
           \right)  \left(w\right)
                   \right]
\nonumber \\
 &  & 
   {}+\left( 1-\frac{3}{4} \alpha \left(\alpha+1\right) \right)
    \oint_{z_{I}} \frac{dw}{2\pi i}
      \frac{2}{\left(z-w\right)^{3}}
      \frac{e^{-\alpha\sigma^{\prime}}}
           {\left(\partial^{2}X^{+}
             \right)^{\frac{3}{4}\alpha\left(\alpha+1\right)}}
         \left(w\right)
\nonumber \\
 & = & 
    \left(1-\frac{3}{4}\alpha\left(\alpha+1\right)\right)
    \oint_{z_{I}}\frac{dw}{2\pi i}
       \frac{2}{\left(z-w\right)^{3}}
       \frac{e^{-\alpha\sigma^{\prime}}}
            {\left(\partial^{2}X^{+}
              \right)^{\frac{3}{4}\alpha\left(\alpha+1\right)}}
         \left(w\right)\ .
\end{eqnarray}
On the assumption that one can replace $X^{+}$
by its expectation value%
\footnote{Here we have also assumed 
          $\partial^{2}\rho\left(z_{I}\right)\ne 0$,
          which is generically true. 
          $\partial^{2}\rho\left(z_{I}\right)=0$
          implies that $z_{I}$ coincides with another interaction 
          point $z_{I^{\prime}}$ $\left(I^{\prime}\ne I\right)$. 
          Such cases are considered as a limit of the generic cases, 
          in which we should insert $\mathcal{O}_{I}$ and 
          $\mathcal{O}_{I^{\prime}}$ at the same point.} 
\begin{eqnarray}
T\left(z\right)\mathcal{O}_{I} 
& \sim & 
  \left( 1-\frac{3}{4}\alpha\left(\alpha+1\right) \right)
  \oint_{z_{I}} \frac{dw}{2\pi i} 
    \frac{2}{\left(z-w\right)^{3}}
    \frac{e^{-\alpha\sigma^{\prime}}}
         {\left( -\frac{i}{2}\partial^{2} \rho
          \right)^{\frac{3}{4}\alpha\left(\alpha+1\right)}}
       \left(w\right)
\nonumber \\
 & \sim & \mbox{regular}\ .
\label{eq:TOI}
\end{eqnarray}
Therefore $\mathcal{O}_{I}$ seems to have the right properties to
be inserted in the free field expression. 
We will prove the fact that the OPE
$T\left(z\right)\mathcal{O}_{I}$ becomes regular
without any assumptions, in the next subsection.

With similar reasonings, one can deduce the singular behaviors of
$b',c'$ at the points $z=Z_{r}$ and $\infty$ as well, from which
one can infer the ghost operators to be inserted at these points.
For $z\sim Z_{r}$, we can see that 
$e^{\alpha\sigma^{\prime}}\left(Z_{r}\right)$
should be inserted. 
Combined with the insertion $e^{-ip_{r}^{+}X^{-}}(Z_{r},\bar{Z}_{r})$
and the operator to be introduced in eq.(\ref{eq:insertionX-}) at
$z=Z_{r}$, 
this ghost operator reproduces the correct OPE with the energy-momentum
tensor. For $w\equiv\frac{1}{z}\sim0$, the ghosts should behave as
\begin{equation}
b^{\prime}\left(w\right)\sim w^{-3}\ ,
 \qquad 
 c^{\prime}\left(w\right)\sim w^{3}\ ,
\end{equation}
and one can see that $e^{3\sigma^{\prime}}\left(\infty\right)$,
which is of weight $-3\alpha$, should be inserted. We can define
a conformal invariant combination, 
\begin{equation}
\mathcal{R} \equiv
  \left| \oint_{\infty} \frac{dz}{2\pi i}
         \partial\Phi\left(\partial X^{+}\right)^{3\alpha}
         e^{3\sigma^{\prime}}\right|^{2}\ ,
\end{equation}
to implement such an insertion.

We should also take care of the singular behavior of $X^{-}$ in the $X^{\pm}$
CFT. {}From the results of Ref.~\cite{Baba:2009ns}, one can see
that $X^{-}$ possesses logarithmic singularities at $Z_{r}$ and
$z_{I^{\left(r\right)}}$, where $z_{I^{(r)}}$ is the interaction
point at which the $r$th string interacts. Therefore it is necessary to  
insert
\begin{equation}
\exp \left( \frac{d-26}{24} \frac{i}{p_{r}^{+}} X^{+} \right)
    \left(Z_{r,}\bar{Z}_{r}\right)\ ,
\qquad
\exp \left( -\frac{d-26}{24} \frac{i}{p_{r}^{+}} X^{+} \right)
     \left(z_{I^{(r)}},\bar{z}_{I^{(r)}}\right)\ .
\label{eq:insertionX-}
\end{equation}
The latter should be made into a conformal invariant combination
\begin{equation}
\mathcal{S}_{r}
  \equiv 
   \oint_{z_{I^{(r)}}} \frac{dz}{2\pi i} \partial\Phi
   \oint_{\bar{z}_{I^{(r)}}} \frac{d\bar{z}}{2\pi i}
       \bar{\partial} \Phi
   \, \exp\left(-\frac{d-26}{24}\frac{i}{p_{r}^{+}}X^{+}\right).
\label{eq:mathcalSr}
\end{equation}

The conformal invariance of $\mathcal{R}$ and $\mathcal{S}_{r}$
can be proved in a similar way to that of $\mathcal{O}_{I}$
in eq.(\ref{eq:TOI})
on the assumption that $X^{+}$ can be replaced by its
expectation value $-\frac{i}{2}(\rho + \bar{\rho})$.
In the next subsection, we will prove that this assumption
is not necessary as in the $\mathcal{O}_{I}$ case.

\subsection{Correlation functions in terms of the free variables}

We have shown what kind of operator insertions are necessary. We would
like to show that they are actually enough and the correlation functions
of the system can be expressed in terms of the free variables only
with the insertions obtained above. To be precise, we will prove 
\begin{eqnarray}
\lefteqn{
  \left\langle 
    \left|e^{3\sigma}\left(\infty\right)\right|^{2}
    \phi_{1} \phi_{2} \cdots \phi_{n}
    \prod_{r=1}^{N}e^{-ip_{r}^{+}X^{-}}\left(Z_{r},\bar{Z}_{r}\right)
  \right\rangle _{X^{\pm},b,c}
} \nonumber \\
& = & 
  \mathcal{C} \left\langle 
      \mathcal{R} \phi_{1} \phi_{2} \cdots \phi_{n}
      \prod_{I} \mathcal{O}_{I}
      \prod_{r=1}^{N} \left[
         \mathcal{S}_{r} \left|\alpha_{r}\right|^{-3\alpha}
          \left| e^{\alpha\sigma'} (Z_{r}) \right|^{2}
         e^{-ip_{r}^{+}X^{\prime-}
            +\frac{d-26}{24}\frac{i}{p_{r}^{+}}X^{+}}
                \left(Z_{r},\bar{Z_{r}}\right)
                      \right]
     \right\rangle _{\mathrm{free}},
\nonumber \\
&  & \label{eq:general}
\end{eqnarray}
where $\mathcal{C}$ is a numerical constant.

Let us first consider the simplest case and check if 
\begin{eqnarray}
\lefteqn{
   \left\langle 
      \left|e^{3\sigma}(\infty)\right|^{2}
      \prod_{r=1}^{N}e^{-ip_{r}^{+}X^{-}}(Z_{r},\bar{Z}_{r})
  \right\rangle _{X^{\pm},b,c}
} \nonumber \\
& \propto & 
    \left\langle
         \mathcal{R}  \prod_{I} \mathcal{O}_{I}
         \prod_{r=1}^{N} \left[
             \mathcal{S}_{r} \left|\alpha_{r}\right|^{-3\alpha}
             \left| e^{\alpha\sigma^{\prime}}(Z_{r}) \right|^{2}
             e^{-ip_{r}^{+}X^{\prime-}
                +\frac{d-26}{24}\frac{i}{p_{r}^{+}}X^{+}}
                    \left(Z_{r},\bar{Z_{r}}\right)
                          \right]
     \right\rangle _{\mathrm{free}}.
\label{eq:correlationfree}
\end{eqnarray}
The left hand side was evaluated to be 
$\exp\left(-\frac{d-26}{24}\Gamma\right)$
up to a constant multiplicative factor in Ref.~\cite{Baba:2009ns}, 
where $\Gamma$ is defined as 
\begin{equation}
e^{-\Gamma}
  =\left|\sum_{r=1}^{N}\alpha_{r}Z_{r}\right|^{4}
   \prod_{r=1}^{N}
        \left( \left|\alpha_{r}\right|^{-2}
               e^{-2\mathop{\mathrm{Re}}\bar{N}_{00}^{rr}}
        \right)
   \prod_{I=1}^{N-2}
         \left|\partial^{2}\rho(z_{I})\right|^{-1}~.
\label{eq:Gamma}
\end{equation}
Here $\bar{N}_{00}^{rr}$ is a Neumann coefficient given by 
\begin{equation}
\bar{N}_{00}^{rr}
  =\frac{\rho(z_{I^{(r)}})}{\alpha_{r}}
   -\sum_{s\neq r}\frac{\alpha_{s}}{\alpha_{r}} \ln(Z_{r}-Z_{s})~.
\end{equation}
It is easy to calculate the free field correlation function 
on the right hand side, and we find that this also becomes 
$\exp\left(-\frac{d-26}{24}\Gamma\right)$
up to a constant multiplicative factor, using the identity, 
\begin{equation}
\frac{\prod_{r>s}\left|Z_{r}-Z_{s} \right|^{2}
      \prod_{I>J}\left|z_{I}-z_{J}\right|^{2}}
     {\prod_{r=1}^{N}\prod_{I}\left|Z_{r}-z_{I}\right|^{2}}
= \left| \sum_{r=1}^{N}\alpha_{r}Z_{r} \right|^{2}
  \frac{\prod_{I}\left|\partial^{2}\rho\left(z_{I}\right)\right|}
       {\prod_{r=1}^{N}\left|\alpha_{r}\right|}\ .
\end{equation}
Thus eq.(\ref{eq:correlationfree}) holds.

Let us then consider the next simplest case where all the $\phi_{i}$
do not include derivatives of $X^{-}$. In this case, $X^{+}$ in
the correlation function can be replaced by its expectation value
$-\frac{i}{2}\left(\rho+\bar{\rho}\right)$. Therefore the problem
is reduced to the case where $\phi_{i}$ are made of ghost fields.
Since the operator insertions on the right hand side was fixed so
that the ghost variables $b,c,\tilde{b},\tilde{c}$ have the same
singularity structure as the quantity on the left hand side, it is
easy to see 
\begin{eqnarray}
\lefteqn{
  \left\langle \mathcal{R} 
      b\left(z_{1}\right) \cdots b\left(z_{n}\right)
      c\left(w_{1}\right)\cdots c\left(w_{n}\right)
      \tilde{b}\left(\bar{u}_{1}\right) \cdots
         \tilde{b}\left(\bar{u}_{m}\right)
      \tilde{c}\left(\bar{v}_{1}\right) \cdots
         \tilde{c}\left(\bar{v}_{m}\right)
      \prod_{I}\mathcal{O}_{I}
   \right.
}\nonumber \\
\lefteqn{
  \quad\times \left.
    \prod_{r=1}^{N}
       \left[  \mathcal{S}_{r}  \left|\alpha_{r}\right|^{-3\alpha}
               \left| e^{\alpha\sigma^{\prime}}(Z_{r}) \right|^{2}
               e^{-ip_{r}^{+}X^{\prime-}
                  +\frac{d-26}{24}\frac{i}{p_{r}^{+}}X^{+}}
                        \left(Z_{r},\bar{Z_{r}}\right)
        \right]
    \right\rangle _{\mathrm{free}}
} \nonumber \\
& \propto & 
   \left\langle \left|e^{3\sigma}\left(\infty\right)\right|^{2}
       b\left(z_{1}\right) \cdots b\left(z_{n}\right)
       c\left(w_{1}\right) \cdots c\left(w_{n}\right)
       \tilde{b}\left(\bar{u}_{1}\right) \cdots
            \tilde{b}\left(\bar{u}_{m}\right)
       \tilde{c}\left(\bar{v}_{1}\right) \cdots
            \tilde{c}\left(\bar{v}_{m}\right)
   \right\rangle _{b,c}
\nonumber \\
&  & \times\left\langle \mathcal{R}
         \prod_{I}\mathcal{O}_{I}
         \prod_{r=1}^{N}
            \left[ \mathcal{S}_{r}  \left|\alpha_{r}\right|^{-3\alpha}
                   \left| e^{\alpha\sigma^{\prime}}(Z_{r}) \right|^{2}
                   e^{-ip_{r}^{+}X^{\prime-}
                      +\frac{d-26}{24}\frac{i}{p_{r}^{+}}X^{+}}
                        \left(Z_{r},\bar{Z}_{r}\right)
             \right]
        \right\rangle _{\mathrm{free}}.
\end{eqnarray}
On the other hand, we have 
\begin{eqnarray}
&  & \left\langle \left| e^{3\sigma}\left(\infty\right) \right|^{2}
         b\left(z_{1}\right) \cdots b\left(z_{n}\right)
         c\left(w_{1}\right) \cdots c\left(w_{n}\right)
\right.\nonumber \\
&  & \qquad\left.
  \times \tilde{b}\left(\bar{u}_{1}\right) \cdots
             \tilde{b}\left(\bar{u}_{m}\right)
         \tilde{c}\left(\bar{v}_{1}\right) \cdots
             \tilde{c}\left(\bar{v}_{m}\right)
         \prod_{r=1}^{N}
              e^{-ip_{r}^{+}X^{-}}\left(Z_{r},\bar{Z}_{r}\right)
\right\rangle _{X^{\pm},b,c}
\nonumber \\
&  & \quad \propto
  \left\langle \left| e^{3\sigma}\left(\infty\right) \right|^{2}
   b\left(z_{1}\right) \cdots b\left(z_{n}\right)
   c\left(w_{1}\right) \cdots c\left(w_{n}\right)
   \tilde{b}\left(\bar{u}_{1}\right) \cdots
       \tilde{b}\left(\bar{u}_{m}\right)
   \tilde{c}\left(\bar{v}_{1}\right) \cdots
       \tilde{c}\left(\bar{v}_{m}\right)
  \right\rangle _{b,c}
\nonumber \\
&  & \hphantom{\qquad\propto\quad}
   \times \left\langle 
             \left|e^{3\sigma}\left(\infty\right)\right|^{2}
             \prod_{r=1}^{N} e^{-ip_{r}^{+}X^{-}}
                               \left(Z_{r},\bar{Z}_{r}\right)
          \right\rangle _{X^{\pm},b,c}.
\label{eq:int-gen-corr}
\end{eqnarray}
Using eq.(\ref{eq:correlationfree}) we can show that these two are
proportional to each other.

\subsubsection*{$X^{-}$ insertions}

Now let us turn to the cases where derivatives of $X^-$ are included in $\phi_i$. 
Once eq.(\ref{eq:general}) is proved for $\phi_{i}$ made of the
derivatives of $X^{+}$ and the ghosts, one can get the free field
expression of the correlation functions with $X^{-}$ insertions by
differentiating eq.(\ref{eq:general}) with respect to 
$p_{r}^{+}$~\cite{Baba:2009ns}.
As an example, let us consider the correlation function 
with one insertion of $\partial X^{-}$
\begin{equation}
 \left\langle 
    \left|e^{3\sigma}\left(\infty\right)\right|^{2}
    \partial X^{-}\left(z\right)
    \prod_{r=1}^{N} e^{-ip_{r}^{+}X^{-}}\left(Z_{r},\bar{Z}_{r}\right)
 \right\rangle _{X^{\pm},b,c}\ .
\end{equation}
It can be expressed in terms of the one with no insertions. 
One can show 
\begin{eqnarray}
\lefteqn{
   \left\langle \left|e^{3\sigma}\left(\infty\right)\right|^{2}
    \partial X^{-}\left(Z_{0}\right)
    \prod_{r=1}^{N}
     \left[  \left|\alpha_{r}\right|^{\frac{d-26}{8}}
             e^{-ip_{r}^{+}X^{-}}\left(Z_{r},\bar{Z}_{r}\right)
     \right]
   \right\rangle _{X^{\pm},b,c}}
\nonumber \\
&  & \propto
 \left. 2i \partial_{Z_{0}} \partial_{\alpha_{0}}
  \left\langle 
      \left|e^{3\sigma}\left(\infty\right)\right|^{2}
      \prod_{r=0}^{N+1}
         \left[ \left|\alpha_{r}\right|^{\frac{d-26}{8}}
                e^{-ip_{r}^{+}X^{-}}\left(Z_{r},\bar{Z}_{r}\right)
         \right]
  \right\rangle _{X^{\pm},b,c}
 \right|_{\alpha_{0}=0}\ .
\end{eqnarray}
Here the factors $\left|\alpha_{r}\right|^{\frac{d-26}{8}}$ are
included so that the limit $\alpha_{0}\to0$ becomes smooth. 
On the right hand side, we have inserted sources 
$e^{-ip_{0}^{+}X^{-}}(Z_{0},\bar{Z}_{0})$
and $e^{-ip_{N+1}^{+}X^{-}}(Z_{N+1},\bar{Z}_{N+1})$ with 
$p_{N+1}^{+}=-p_{0}^{+}=-\frac{1}{2}\alpha_{0}$,
to generate the insertions of $X^{-}$. 
With these sources, $X^{+}$ has the expectation value 
$-\frac{i}{2}\left(\hat{\rho}(z)+\bar{\hat{\rho}}(\bar{z})\right)$,
where 
$\hat{\rho}(z)\equiv\sum_{r=0}^{N+1}\alpha_{r}\ln\left(z-Z_{r}\right)$.
$\hat{\rho}(z)$ has $N$ interaction points. 
In the limit $\alpha_{0}\rightarrow0$,
two of them, which we denote by 
   $\hat{z}_{I^{(0)}}$ and $\hat{z}_{I^{(N+1)}}$,
tend to $Z_{0}$ and $Z_{N+1}$, 
and the other $\hat{z}_{I}$'s go
to the interaction points $z_{I}$ of $\rho(z)$, which are denoted
with the same subscripts~\cite{Baba:2009ns}.

Now let us rewrite the expression on the right hand side using the
free variables. 
By making use of eq.(\ref{eq:correlationfree}) with
$\rho$ replaced by $\hat{\rho}$, we have 
\begin{eqnarray}
 &  & \left. 2i\partial_{Z_{0}} \partial_{\alpha_{0}}
        \left\langle \left|e^{3\sigma}\left(\infty\right)\right|^{2}
          \prod_{r=0}^{N+1}
            \left[ \left|\alpha_{r}\right|^{\frac{d-26}{8}}
                   e^{-ip_{r}^{+}X^{-}}\left(Z_{r},\bar{Z}_{r}\right)
            \right]
        \right\rangle _{X^{\pm},b,c}
      \right|_{\alpha_{0}=0}
\nonumber \\
&  & 
  \ = 2i\partial_{Z_{0}} \partial_{\alpha_{0}}
       \left\langle \mathcal{R}
           \prod_{I=1}^{N-2} \hat{\mathcal{O}}_{I}
           \hat{\mathcal{O}}_{I^{(0)}} \hat{\mathcal{O}}_{I^{(N+1)}}
      \right.
\nonumber \\
&  & \hphantom{=2i\partial_{Z_{0}}\partial_{\alpha_{0}}}
  \left. \left. 
     \times 
      \prod_{r=0}^{N+1}
        \left[ \hat{\mathcal{S}}_{r}
               \left|\alpha_{r}
                \right|^{\frac{3}{2}\alpha\left(\alpha+1\right)}
               \left|e^{\alpha\sigma^{\prime}}(Z_{r})\right|^{2}
               e^{-ip_{r}^{+}X^{\prime-}
                  +\frac{d-26}{24}\frac{i}{p_{r}^{+}}X^{+}}
                      (Z_{r},\bar{Z_{r}})
        \right]
   \right\rangle _{\mathrm{free}}
   \right|_{\alpha_{0}=0},~~~~~~~
\label{eq:onepoint}
\end{eqnarray}
where $\hat{\mathcal{O}}_{I}$ and $\hat{\mathcal{S}}_{r}$ 
are respectively $\mathcal{O}_{I}$ and $\mathcal{S}_{r}$ 
with $z_{I}$ being replaced by $\hat{z}_{I}$. 
Since $\hat{z}_{I^{\left(0\right)}}\to Z_{0}$
and $\hat{z}_{I^{\left(N+1\right)}}\to Z_{N+1}$ 
in the limit $\alpha_{0}\to0$
which we should eventually take, 
we get some operator insertions at
$z=Z_{0}$ and $z=Z_{N+1}$ as a result. 
These insertions should correspond to 
$X^{-}\left(Z_{0}\right)-X^{-}\left(Z_{N+1}\right)$. 
The behaviors of $\hat{z}_{I^{\left(0\right)}}-Z_{0}$, 
$\mathop{\mathrm{Re}}\hat{\bar{N}}_{00}^{00}$
and 
$\partial^{2}\hat{\rho}\left(\hat{z}_{I^{\left(0\right)}}\right)$
in the limit $\alpha_{0}\to0$ are given by 
eqs.(D.1), (D.2) and (D.4) of Ref.~\cite{Baba:2009ns} respectively, 
where $\hat{\bar{N}}_{00}^{00}$
is a Neumann coefficient corresponding to $\hat{\rho}(z)$. 
In the free field correlation function on the right hand side 
of eq.(\ref{eq:onepoint}),
$X^{\prime-}$'s appear only in the form of the vertex operator 
$e^{-ip^{+}X^{\prime-}}$.
Therefore one can replace $X^{+}$ by its expectation value and vice
versa. Using these facts, we can show that
in the limit $\alpha_0\to 0$,
\begin{eqnarray}
\lefteqn{
  \hat{\mathcal{O}}_{I^{\left(0\right)}}
  \hat{\mathcal{S}}_{0}
  \left|\alpha_{0}\right|^{\frac{3}{2}\alpha\left(\alpha+1\right)}
  \left| e^{\alpha\sigma^{\prime}}(Z_{0}) \right|^{2}
  e^{-ip_{0}^{+}X^{\prime-}+\frac{d-26}{24}\frac{i}{p_{0}^{+}}X^{+}}
        \left(Z_{0},\bar{Z_{0}}\right)
}  \nonumber \\
& \sim & 
  \left|\alpha_{0}\right|^{\frac{3}{2}\alpha\left(\alpha+1\right)}
  \left|\partial^{2}\hat{\rho}\left(\hat{z}_{I^{\left(0\right)}}\right)
   \right|^{-\frac{3}{2}\alpha\left(\alpha+1\right)}
  \left|e^{-\alpha\sigma^{\prime}}
           \left(\hat{z}_{I^{\left(0\right)}}\right)
  \right|^{2}
\nonumber \\
&  & \quad \times
  \left| e^{\alpha\sigma^{\prime}}(Z_{0})\right|^{2}
  e^{-ip_{0}^{+}X^{\prime-}}\left(Z_{0},\bar{Z_{0}}\right)
  e^{-\frac{d-26}{12}\mathop{\mathrm{Re}}\hat{\bar{N}}_{00}^{00}}
\nonumber \\
 & \sim & 
  1 - ip_{0}^{+}X^{\prime-}\left(Z_{0},\bar{Z}_{0}\right)
    + \alpha_{0} \alpha
       \left( \frac{\partial\sigma^{\prime}}{\partial\rho}
                   \left(Z_{0}\right)
              +\frac{\bar{\partial}\tilde{\sigma}^{\prime}}
                    {\bar{\partial}\bar{\rho}} 
                   \left(\bar{Z}_{0}\right)
       \right)
    + \left(2\alpha^{2}+3\alpha\right)
      \mathop{\mathrm{Re}} \frac{\partial^{2}\rho}
                                {\left(\partial\rho\right)^{2}}
                                  \left(Z_{0}\right)
\nonumber \\
 & \sim & 
    1 - \frac{i}{2} \alpha_{0}
         \left[ X^{\prime-}
                +\alpha \left( 
                    \frac{\partial\sigma^{\prime}}{\partial X^{+}}
                    + \frac{\bar{\partial}\tilde{\sigma}^{\prime}}
                           {\bar{\partial}X^{+}}  \right)
                +\left( \alpha^{2} +\frac{3}{2}\alpha \right)
                   \left(\frac{\partial^{2}X^{+}}
                              {\left(\partial X^{+}\right)^{2}}
                        +\frac{\bar{\partial}^{2}X^{+}}
                              {\left(\bar{\partial}X^{+}\right)^{2}}
                   \right)
          \right]  (Z_{0},\bar{Z}_{0})
\nonumber \\
& = & 1-\frac{i}{2}\alpha_{0}X^{-}\left(Z_{0},\bar{Z}_{0}\right)\ ,
\label{eq:X-0}
\end{eqnarray}
and similarly 
\begin{eqnarray}
&  & \hat{\mathcal{O}}_{I^{\left(N+1\right)}}
     \hat{\mathcal{S}}_{N+1}
     \left| \alpha_{0} \right|^{\frac{3}{2}\alpha\left(\alpha+1\right)}
     \left| e^{\alpha\sigma^{\prime}}(Z_{N+1}) \right|^{2}
     e^{ip_{0}^{+}X^{\prime-}-\frac{d-26}{24}\frac{i}{p_{0}^{+}}X^{+}}
          \left(Z_{N+1},\bar{Z}_{N+1}\right)
\nonumber \\
 &  & \qquad \sim
   1+\frac{i}{2}\alpha_{0}X^{-}\left(Z_{N+1},\bar{Z}_{N+1}\right)\ .
\label{eq:X-N+1}
\end{eqnarray}
Substituting eqs.(\ref{eq:X-0}) and (\ref{eq:X-N+1}) 
into eq.(\ref{eq:onepoint}), we 
obtain\footnote{Here (and in eqs.(\ref{eq:general})(\ref{eq:twoX-})) 
                $z_{I}$ which appears in the definitions of 
                $\mathcal{O}_{I},\mathcal{S}_{r}$ are taken to be 
                the interaction point which correspond to 
                the Mandelstam mapping 
                  $\rho\left(z\right)
                     =\sum_{r=1}^{N}\alpha_{r}\ln\left(z-Z_{r}\right)$.}
\begin{eqnarray}
\lefteqn{
   \left\langle \left|e^{3\sigma}\left(\infty\right)\right|^{2}
       \partial X^{-}(Z_{0})
       \prod_{r=1}^{N} e^{-ip_{r}^{+}X^{-}} \left(Z_{r},\bar{Z}_{r}\right)
   \right\rangle _{X^{\pm}b,c}
}
\label{eq:oneinsertion}\\
 & \propto & 
   \left\langle \mathcal{R}\partial X^{-}(Z_{0})
        \prod_{I=1}^{N-2} \mathcal{O}_{I}
        \prod_{r=1}^{N}
            \left[ \mathcal{S}_{r} 
                   \left|\alpha_{r}\right|^{-3\alpha}
                   \left|e^{\alpha\sigma^{\prime}}(Z_{r})\right|^{2}
                   e^{-ip_{r}^{+}X^{\prime-}
                      +\frac{d-26}{24}\frac{i}{p_{r}^{+}}X^{+}}
                         \left(Z_{r},\bar{Z_{r}}\right)
            \right]
    \right\rangle _{\mathrm{free}}.
\nonumber 
\end{eqnarray}

It is straightforward to prove eq.(\ref{eq:general}) for more general
insertions. The key relation is eq.(\ref{eq:X-0}), which is valid
in the free field correlation functions in which $X^{\prime-}$'s
appear only in the form of the vertex operator $e^{-ip^{+}X^{\prime-}}$.
For example, let us consider the correlation functions with two insertions
of $\partial X^{-}$ 
\begin{eqnarray}
\lefteqn{
   \left\langle \left|e^{3\sigma}\left(\infty\right)\right|^{2}
        \partial X^{-}\left(z\right)
        \partial X^{-}\left(Z_{0}\right)
       \prod_{r=1}^{N}
          \left[ \left|\alpha_{r}\right|^{\frac{d-26}{8}}
                 e^{-ip_{r}^{+}X^{-}}\left(Z_{r},\bar{Z}_{r}\right)
          \right]
  \right\rangle _{X^{\pm},b,c}
}\nonumber \\
 &  & \propto \left.
   2i \partial_{Z_{0}} \partial_{\alpha_{0}}
     \left\langle \left|e^{3\sigma}\left(\infty\right)\right|^{2}
         \partial X^{-}\left(z\right)
         \prod_{r=0}^{N+1}
            \left[ \left|\alpha_{r}\right|^{\frac{d-26}{8}}
                   e^{-ip_{r}^{+}X^{-}}\left(Z_{r},\bar{Z}_{r}\right)
            \right]
  \right\rangle _{X^{\pm},b,c}\right|_{\alpha_{0}=0}~.~~~~~
\end{eqnarray}
Using eq.(\ref{eq:oneinsertion}), 
the right hand side is expressed as 
\begin{eqnarray}
 &  & 2i\partial_{Z_{0}}\partial_{\alpha_{0}}
   \left\langle \mathcal{R}
       \prod_{I=1}^{N-2} \hat{\mathcal{O}}_{I}
         \hat{\mathcal{O}}_{I^{(0)}} \hat{\mathcal{O}}_{I^{(N+1)}}
       \partial X^{-}\left(z\right)
\right. \nonumber \\
 &  & \hphantom{2i\partial_{Z_{0}}\partial_{\alpha_{0}}\langle\;}
   \left.\left. \times
     \prod_{r=0}^{N+1}
       \left[ \hat{\mathcal{S}}_{r}
              \left|\alpha_{r}
               \right|^{\frac{3}{2}\alpha\left(\alpha+1\right)}
              \left| e^{\alpha\sigma^{\prime}}(Z_{r}) \right|^{2}
              e^{-ip_{r}^{+}X^{\prime-}
                 +\frac{d-26}{24}\frac{i}{p_{r}^{+}}X^{+}}
                     (Z_{r},\bar{Z_{r}})
        \right]
    \right\rangle _{\mathrm{free}}
    \right|_{\alpha_{0}=0}.~~~~~~~
\label{eq:2X-free}
\end{eqnarray}
Here we would like to use eq.(\ref{eq:X-0}) to deal with the limit
$\alpha_{0}\to0$. In this form, eq.(\ref{eq:X-0}) does not hold apparently,
because of the presence of $\partial X^{-}\left(z\right)$. 
However,
$\partial X^{-}\left(z\right)$ can be rewritten as 
\begin{eqnarray}
\partial X^{-}\left(z\right) 
& = & \partial X^{\prime-}\left(z\right)+\cdots\nonumber \\
& = & \left.i \partial_{z}\partial_{p^{+}}
                 e^{-ip^{+}X^{\prime-}}\left(z\right)
     \right|_{p^{+}=0}+\cdots\ ,
\label{eq:delX-}
\end{eqnarray}
where $\cdots$ denotes the quantities which does not involve $X^{\prime-}$.
Substituting this into eq.(\ref{eq:2X-free}), we can make it into
the form where $X^{\prime-}$ is exponentiated so that eq.(\ref{eq:X-0})
holds. Thus we can show 
\begin{eqnarray}
\lefteqn{
  \left\langle \left|e^{3\sigma}\left(\infty\right)\right|^{2}
   \partial X^{-}\left(z\right)\partial X^{-}(Z_{0})
   \prod_{r=1}^{N} e^{-ip_{r}^{+}X^{-}}\left(Z_{r},\bar{Z}_{r}\right)
  \right\rangle _{X^{\pm}b,c}}
\nonumber \\
& \propto & 
  \left\langle  \mathcal{R} \partial X^{-} \left(z\right)
      \partial X^{-}(Z_{0})
      \prod_{I=1}^{N-2} \mathcal{O}_{I}
  \right.
\nonumber \\
&  & \left. \hphantom{\langle\ }
  \times \prod_{r=1}^{N}
            \left[ \mathcal{S}_{r}
                   \left|\alpha_{r}\right|^{-3\alpha}
                   \left|e^{\alpha\sigma^{\prime}}(Z_{r})\right|^{2}
                   e^{-ip_{r}^{+}X^{\prime-}
                      +\frac{d-26}{24}\frac{i}{p_{r}^{+}}X^{+}}
                          \left(Z_{r},\bar{Z_{r}}\right)
            \right]
\right\rangle _{\mathrm{free}}.
\label{eq:twoX-}
\end{eqnarray}
Proceeding in this way, we can show that eq.(\ref{eq:general}) holds
for general $\phi_{i}$ and the correlation functions can be expressed
by using the free variables.

{}From eq.(\ref{eq:general}) one can see that 
$\mathcal{O}_{I}$, $\mathcal{R}$, $\mathcal{S}_r$
are conformal invariant, which has been proved 
in the previous subsection
assuming that $X^{+}$ can be replaced by its expectation value.
Indeed, when one of $\phi_i$ is the energy-momentum tensor $T(z)$, 
the left hand side is not singular in the limit $z\to z_I,\infty$. 
On the right hand side, this fact implies that
 $\mathcal{O}_{I}$, $\mathcal{R}$, $\mathcal{S}_r$
are conformal invariant.

\section{Free variables: supersymmetric case 
         \label{sec:supersymmetric-case}}

Supersymmetric case can be dealt with in a similar way, using the
superspace formulation. In this section, we denote some superfields
by using the same symbols as those in the bosonic case. We think that
this does not cause any confusion.

\subsection{Supersymmetric $X^{\pm}$ CFT}

Supersymmetric $X^{\pm}$ CFT can be defined by using the superspace
formalism. It is described by the superfield variables
\begin{equation}
\mathcal{X}^{\pm}\left(\mathbf{z},\bar{\mathbf{z}}\right)
  \equiv X^{\pm}\left(z,\bar{z}\right)
         +i\theta\psi^{\pm}\left(z\right)
         +i\theta\tilde{\psi}^{\pm}\left(\bar{z}\right)
         +i\theta\bar{\theta}F^{\pm}\left(z,\bar{z}\right)\ ,
\end{equation}
where $\mathbf{z}=\left(z,\theta\right)$ is the superspace coordinate.
The energy-momentum tensor is given as~\cite{Baba:2009fi}
\begin{equation}
\frac{1}{2}D\mathcal{X}^{+}\partial\mathcal{X}^{-}
  +\frac{1}{2}D\mathcal{X}^{-}\partial\mathcal{X}^{+}
  -\frac{d-10}{4}S(\mathbf{z},\mathbf{X}_{L}^{+})~.
\label{eq:sXpmem}
\end{equation}
Here $S(\mathbf{z},\mathbf{X}_{L}^{+})$ is the super Schwarzian
derivative,
\begin{equation}
S(\mathbf{z},\mathbf{X}_{L}^{+}) 
 \equiv \frac{D^{4}\Theta^{+}}{D\Theta^{+}}
         -2\frac{D^{3}\Theta^{+}D^{2}\Theta^{+}}{(D\Theta^{+})^{2}}\\
  =  -\frac{1}{4}D\Phi\partial\Phi+\frac{1}{2}\partial D\Phi~.
\end{equation}
The superspace coordinate 
$\mathbf{X}_{L}^{+}$ is defined as 
$\mathbf{X}_{L}^{+}\equiv\left(\mathcal{X}_{L}^{+},\Theta^{+}\right)$,
where
$\mathcal{X}_{L}^{+}$ denotes the left-moving part of $\mathcal{X}^{+}$
and
\begin{equation}
\Theta^{+}(\mathbf{z}) 
  \equiv \frac{D\mathcal{X}^{+}}
              {\left(\partial\mathcal{X}^{+}\right)^{\frac{1}{2}}}
                (\mathbf{z})~,
\qquad
\Phi\left(\mathbf{z,\bar{\mathbf{z}}}\right) 
   \equiv 
   \ln \left( -4\left(D\Theta^{+}\right)^{2}
                \left(\bar{D}\tilde{\Theta}^{+}\right)^{2}\right)~.
\end{equation}

Similarly to the bosonic case, we consider the correlation functions
with the insertion 
 $\prod_{r=1}^{N} e^{-p_{r}^{+}\mathcal{X}^{-}}
                   \left(\mathbf{Z}_{r},\bar{\mathbf{Z}}_{r}\right)$.
With this insertion, 
$\mathcal{X}^{+}\left(\mathbf{z},\bar{\mathbf{z}}\right)$
has an expectation value
 $-\frac{i}{2}
   \left(\rho\left(\mathbf{z}\right)
         +\bar{\rho}\left(\mathbf{z}\right)\right)$
where 
 $\rho\left(\mathbf{z}\right)
   \equiv
   \sum_{r=1}^{N}\alpha_{r}\ln\left(\mathbf{z}-\mathbf{Z}_{r}\right)$
is the super Mandelstam mapping~\cite{Baba:2009fi}.

The variables $\mathcal{X}^{\pm}$ satisfy the OPE's
\begin{eqnarray}
D\mathcal{X}^{+}\left(\mathbf{z}\right)
 D\mathcal{X}^{+}\left(\mathbf{z}^{\prime}\right) 
& \sim & \mathrm{regular}\ ,
\nonumber \\
D\mathcal{X}^{-}\left(\mathbf{z}\right)
 D\mathcal{X}^{+}\left(\mathbf{z}^{\prime}\right) 
& \sim & \frac{1}{\mathbf{z}-\mathbf{z}^{\prime}}\ ,
\nonumber \\
D\mathcal{X}^{-}\left(\mathbf{z}\right)
 D\mathcal{X}^{-}\left(\mathbf{z}^{\prime}\right) 
& \sim & -\frac{d-10}{4}DD^{\prime}
    \left[ \frac{3\Theta^{+}\left(\mathbf{z}\right)
                  \Theta^{+}\left(\mathbf{z}^{\prime}\right)}
                {\left(\mathcal{X}_{L}^{+}\left(\mathbf{z}\right)
                   -\mathcal{X}_{L}^{+}\left(\mathbf{z}^{\prime}\right)
                   -\Theta^{+}\left(\mathbf{z}\right)
                     \Theta^{+}\left(\mathbf{z}^{\prime}\right)
                 \right)^{3}}
    \right. \nonumber \\
 &  & \hphantom{-\frac{d-10}{4}DD^{\prime}\ }
    \left. 
      {}+\frac{1}{2\left(\mathcal{X}_{L}^{+}\left(\mathbf{z}\right)
                   -\mathcal{X}_{L}^{+}\left(\mathbf{z}^{\prime}\right)
                   -\Theta^{+}\left(\mathbf{z}\right)
                    \Theta^{+}\left(\mathbf{z}^{\prime}\right)
                     \right)^{2}}
     \right].~~~
\end{eqnarray}
 The right hand side of the third equation should be treated as in
the bosonic case and we get the form of the OPE in Ref.\ \cite{Baba:2009fi}.
Using these OPE's, one can show that the energy-momentum tensor in
eq.(\ref{eq:sXpmem}) satisfies the super Virasoro algebra with $\hat{c}=12-d$.
It follows that together with the super-reparametrization ghosts and
the transverse part, the total central charge becomes $0$.

\subsection{Free fields}

As in the bosonic case, we consider the
system which consists of the supersymmetric $X^{\pm}$ CFT and the
super-reparametrization ghosts. The ghost variables are described
by the superfields 
$B\left(\mathbf{z}\right)$, $C\left(\mathbf{z}\right)$,
$\tilde{B}\left(\bar{\mathbf{z}}\right)$, 
$\tilde{C}\left(\mathbf{z}\right)$,
which are given as 
\begin{eqnarray}
 &  & 
  B\left(\mathbf{z}\right)
      \equiv
      \beta\left(z\right)+\theta b\left(z\right)\ ,
\qquad
  \tilde{B}\left(\bar{\mathbf{z}}\right)
    \equiv\tilde{\beta}\left(\bar{z}\right)
          +\bar{\theta}\tilde{b}\left(\bar{z}\right)~,
\nonumber \\
 &  & 
    C\left(\mathbf{z}\right)
     \equiv c\left(z\right) +\theta\gamma\left(z\right)\ ,
\qquad
  \tilde{C}\left(\bar{\mathbf{z}}\right)
     \equiv \tilde{c}\left(\bar{z}\right)
            +\bar{\theta}\tilde{\gamma}\left(\bar{z}\right)\ .
\label{eq:ghost-sf}
\end{eqnarray}
The free superfields $\mathcal{X}^{+}$, $\mathcal{X}^{\prime-}$
and the ghosts $B^{\prime}$, $C^{\prime}$, $\tilde{B}^{\prime}$,
$\tilde{C}^{\prime}$ with weights $\left(\frac{3}{2}+\alpha,0\right)$,
$\left(-1-\alpha,0\right)$, $\left(0,\frac{3}{2}+\alpha\right)$,
$\left(0,-1-\alpha\right)$ can be defined as 
\begin{eqnarray}
&  & 
 \qquad 
   B^{\prime}\left(\mathbf{z}\right)
      \equiv \left(D\Theta^{+}\right)^{2\alpha}
              B\left(\mathbf{z}\right)\ ,
\qquad\;
    \tilde{B}^{\prime}\left(\bar{\mathbf{z}}\right)
      \equiv \left(\bar{D}\tilde{\Theta}^{+}\right)^{2\alpha}
              \tilde{B}\left(\bar{\mathbf{z}}\right)\ ,
\nonumber \\
&  & 
\qquad 
  C^{\prime}\left(\mathbf{z}\right)
   \equiv\left(D\Theta^{+}\right)^{-2\alpha}
          C\left(\mathbf{z}\right)\ ,
\qquad
   \tilde{C}^{\prime}\left(\bar{\mathbf{z}}\right)
      \equiv \left(\bar{D}\tilde{\Theta}^{+}\right)^{-2\alpha}
             \tilde{C}\left(\bar{\mathbf{z}}\right)\ ,
\nonumber \\
 &  & 
\mathcal{X}^{\prime-}\left(\mathbf{z},\bar{\mathbf{z}}\right)
   \equiv
     \mathcal{X}^{-}\left(\mathbf{z},\bar{\mathbf{z}}\right)
\nonumber \\
 &  & \hphantom{\mathcal{X}^{\prime-}
                \left(\mathbf{z},\bar{\mathbf{z}}\right)
                \equiv}
     {}+\alpha \left[
          \partial D \left(\Sigma^{\prime}+\frac{1}{2}\Phi\right)
          \frac{\Theta^{+}}{\left(D\Theta^{+}\right)^{3}}
          -\partial \left(\Sigma^{\prime}+\frac{1}{2}\Phi\right)
           \left(\frac{1}{\left(D\Theta^{+}\right)^{2}}
                  +\frac{\partial\Theta^{+}\Theta^{+}}
                        {\left(D\Theta^{+}\right)^{4}}
           \right)\right.
\nonumber \\
 &  & \hphantom{\mathcal{X}^{\prime-}
                \left(\mathbf{z},\bar{\mathbf{z}}\right)
                \equiv-\alpha]\quad}
   {}-D\left(\Sigma^{\prime}+\frac{1}{2}\Phi\right)
      \left(\frac{\partial\Theta^{+}}{\left(D\Theta^{+}\right)^{3}}
             +\frac{\partial D\Theta^{+}\Theta^{+}}
                    {\left(D\Theta^{+}\right)^{4}}
       \right)
\nonumber \\
 &  & \hphantom{\mathcal{X}^{\prime-}
                \left(\mathbf{z},\bar{\mathbf{z}}\right)
                \equiv-\alpha]\ }
    \left.
    {}+\mathrm{c.c.}
     \vphantom{\frac{\Theta^{+}}{\left(D\Theta^{+}\right)^{3}}}
    \right]~.
\label{eq:Xprime-}
\end{eqnarray}
Here 
\begin{equation}
\alpha=\frac{d-10}{8}~,\label{eq:alpha-super}
\end{equation}
 and 
\begin{equation}
\Sigma^{\prime}\left(\mathbf{z}\right) 
\equiv 
\sigma^{\prime}\left(z\right)-\phi^{\prime}\left(z\right)
  -\theta \beta^{\prime}c^{\prime}\left(z\right)~,
\end{equation}
 where $\sigma^{\prime}$ and $\phi^{\prime}$ are defined so that
$\partial\sigma^{\prime}=c^{\prime}b^{\prime}$ and 
\begin{equation}
\beta^{\prime}(z)=e^{-\phi^{\prime}}\partial\xi^{\prime}(z)~,
\qquad
\gamma^{\prime}(z)=\eta^{\prime}e^{\phi^{\prime}}(z)~.
\label{eq:betaprime-bosonize}
\end{equation}
We note that 
\begin{equation}
CB\left(\mathbf{z}\right)
  =C^{\prime}B^{\prime}\left(\mathbf{z}\right)
  =-D\Sigma^{\prime}\left(\mathbf{z}\right)\ .
\end{equation}
The OPE's between 
$\mathcal{X}^{+}$, $\mathcal{X}^{\prime-}$, $B^{\prime}$,
$C^{\prime}$, $\tilde{B}^{\prime}$, $\tilde{C}^{\prime}$
can be derived from the OPE's of 
$\mathcal{X}^{\pm}$, $B$, $C$, $\tilde{B}$, $\tilde{C}$
and one can see that they are free variables.

The total energy-momentum tensor, 
\begin{eqnarray}
T\left(\mathbf{z}\right)
   &=& \frac{1}{2}D\mathcal{X}^{+}\partial\mathcal{X}^{-}
    +\frac{1}{2}D\mathcal{X}^{-}\partial\mathcal{X}^{+}
    -\frac{d-10}{4}
       S\left(\mathbf{z},
              \mbox{\mathversion{bold} $\mathcal{X}$}_{L}^{+}
         \right)
\nonumber \\
 &  & 
{}+\frac{1}{2}DCDB
  -\frac{3}{2}\partial CB -C\partial B\ ,
\label{eq:EMsuper1}
\end{eqnarray}
can be rewritten in terms of the free fields as 
\begin{eqnarray}
T\left(\mathbf{z}\right) 
& = & 
 \frac{1}{2}D\mathcal{X}^{+}\partial\mathcal{X}^{\prime-}
 +\frac{1}{2}D\mathcal{X}^{\prime-}\partial\mathcal{X}^{+}
\nonumber \\
 &  & 
  {}+\frac{1}{2}DC^{\prime}DB^{\prime}
   -\frac{1}{2}\partial C^{\prime}B^{\prime}
   -\left(1+\alpha\right)\partial\left(C^{\prime}B^{\prime}\right)\ ,
\end{eqnarray}
which is the energy-momentum tensor for the free fields 
$\mathcal{X}^{+}$,
$\mathcal{X}^{\prime-}$, $B^{\prime}$, $C^{\prime}$.
It is also possible to express 
$\mathcal{X}^{+}$,
$\mathcal{X}^{-}$, $B$, $C$, $\tilde{B}$, $\tilde{C}$ in terms of 
the free variables.

\subsection{Operator insertions}

Let us define the correlation functions 
on the complex plane 
$\left\langle \phi_{1}\cdots\phi_{n}
  \right\rangle _{\mathcal{X}^{\pm},B,C}$
and 
$\left\langle \phi_{1}\cdots\phi_{n}
  \right\rangle _{\mathrm{free}}$
as in the bosonic case. 
The correlation functions that we are interested
in are of the form 
\begin{equation}
\left\langle 
  \left| e^{3\sigma-2\phi}(\infty) \right|^{2}
  \phi_{1}\phi_{2}\cdots\phi_{n}
  \prod_{r=1}^{N}e^{-ip_{r}^{+}\mathcal{X}^{-}}
                  \left(\mathbf{Z}_{r},\bar{\mathbf{Z}}_{r}\right)
\right\rangle _{\mathcal{X}^{\pm},B,C}~.
\end{equation}
The ghosts are bosonized in the usual way and 
$\left|e^{3\sigma-2\phi}\left(\infty\right)\right|^{2}$
is inserted to soak up the ghost zero modes.

As in the bosonic case, the correlation functions of 
the superconformal field theory for 
$\mathcal{X}^{\pm}$, $B$, $C$, $\tilde{B}$, $\tilde{C}$
can be expressed as the correlation functions 
of the free field theory
with operator insertions at $\mathbf{z}=\tilde{\mathbf{z}}_{I}$,
$\mathbf{Z}_{r}$ and $\infty$. 
Here $\tilde{\mathbf{z}}_{I}$ $(I=1,\ldots,N-2)$
denote the points determined by 
$\partial\rho(\tilde{\mathbf{z}}_{I})
  =\partial D\rho(\tilde{\mathbf{z}}_{I})=0$
~\cite{Berkovits:1985ji, Berkovits:1987gp, Aoki:1990yn}.

Let us consider the operator which should be inserted at 
$\mathbf{z}=\tilde{\mathbf{z}}_{I}$
to realize the singular behaviors of the ghost fields at this point.
It is a little bit complicated, compared with the bosonic case, but
straightforward to show that in order for the variables 
$B\left(\mathbf{z}\right)$
and $C\left(\mathbf{z}\right)$ to be regular at 
$\mathbf{z}=\tilde{\mathbf{z}}_{I}$,
we need to insert 
\begin{equation}
\left[ 
  1-\alpha\frac{D\rho}{\partial^{2}\rho}\partial D\Sigma^{\prime}
   -\alpha\frac{\partial^{2}D\rho D\rho}
               {\left(\partial^{2}\rho\right)^{2}}
      \partial\Sigma^{\prime}
\right]
e^{-\alpha\Sigma^{\prime}}\left(\tilde{\mathbf{z}}_{I}\right)\ .
\label{eq:ghostpart}
\end{equation}
We should make up a superconformal invariant operator insertions
whose ghost part is eq.(\ref{eq:ghostpart}). 
The form of such insertions can be read off from the 
partition function. 
The partition function
of the $X^{\pm}$ CFT with the insertion 
$\prod_{r=1}^{N} e^{-ip_{r}^{+}\mathcal{X}^{-}}
                    (\mathbf{Z}_{r},\bar{\mathbf{Z}}_{r})$
becomes 
$e^{-\alpha\Gamma_{\mathrm{super}}}$~\cite{Baba:2009fi,%
Berkovits:1985ji,Berkovits:1987gp},
where 
\begin{equation}
e^{-\Gamma_{\mathrm{super}}}
 =|A|^{2}
   \prod_{I=1}^{N-2}
      \left| \left(\partial^{2}\rho
                     -\frac{5}{3}
                      \frac{\partial^{3}D\rho D\rho}{\partial^{2}\rho}
                     +3\frac{\partial^{3}\rho\partial^{2}D\rho D\rho}
                            {\left(\partial^{2}\rho\right)^{2}}
              \right)  (\tilde{\mathbf{z}}_{I})
      \right|^{-\frac{1}{2}}
  \prod_{r=1}^{N}
      \left( \left|\alpha_{r}\right|^{-1}
             e^{-\mathop{\mathrm{Re}}\bar{N}_{00}^{rr}}
      \right)~,
\label{eq:Gammasuper}
\end{equation}
with 
\begin{eqnarray}
A 
 & \equiv & \sum_{r}\alpha_{r}Z_{r}
            -\frac{\sum_{r}\alpha_{r}\Theta_{r}
                     \sum_{r}\alpha_{r}\Theta_{r}Z_{r}}
                   {\sum_{r}\alpha_{r}Z_{r}}\ ,
\nonumber \\
\bar{N}_{00}^{rr} 
  & \equiv & 
    \frac{\rho(\tilde{\mathbf{z}}_{I^{(r)}})}{\alpha_{r}}
     -\sum_{s\neq r} \frac{\alpha_{s}}{\alpha_{r}}
                     \ln\left(\mathbf{Z}_{r}-\mathbf{Z}_{s}\right)~,
\end{eqnarray}
and $\tilde{\mathbf{z}}_{I^{(r)}}$ 
denotes one of $\tilde{\mathbf{z}}_{I}$'s such that
at $\rho = \rho (\tilde{\mathbf{z}}_{I^{(r)}})$
the $r$th string interacts. 
{}From this, one can see that the insertion~(\ref{eq:ghostpart})
should come with 
\begin{equation}
\left(\partial^{2}\rho
      -\frac{5}{3}\frac{\partial^{3}D\rho D\rho}{\partial^{2}\rho}
      +3\frac{\partial^{3}\rho\partial^{2}D\rho D\rho}
             {\left(\partial^{2}\rho\right)^{2}}
\right)^{-\frac{\alpha}{4}}
  \left(\tilde{\mathbf{z}}_{I}\right)\ .
\end{equation}
Indeed, as we will see later,
with the factor so arranged, 
one can obtain eq.(\ref{eq:supercorrelationfree}). 
Therefore we define the following operator 
\begin{eqnarray}
\mathcal{O}_{I} 
  & \equiv & 
     \left| \oint_{\tilde{\mathbf{z}}_{I}} \frac{d\mathbf{z}}{2\pi i}
                D\Phi
       \left[
         1+\frac{\alpha}{12}
           \frac{\partial^{3}D\mathcal{X}^{+}D\mathcal{X}^{+}}
                {\left(\partial^{2}\mathcal{X}^{+}\right)^{2}}
          +\alpha \left(\frac{\alpha}{32}-\frac{1}{8}\right)
           \frac{\partial^{3}\mathcal{X}^{+}\partial^{2}
                  D\mathcal{X}^{+}D\mathcal{X}^{+}}
                {\left(\partial^{2}\mathcal{X}^{+}\right)^{3}}
\right.\right.\nonumber \\
&  & \hphantom{\oint_{\tilde{\mathbf{z}}_{I}}
                \frac{d\mathbf{z}}{2\pi i}
                D\Phi\qquad}
 {}-\frac{\alpha^{2}}{8}
          \frac{\partial^{3}\mathcal{X}^{+}D\mathcal{X}^{+}}
               {\left(\partial^{2}\mathcal{X}^{+}\right)^{2}}
          D\Sigma^{\prime}
           +\frac{\alpha^{2}}{8}
            \frac{\partial^{2}D\mathcal{X}^{+}D\mathcal{X}^{+}}
                 {\left(\partial^{2}\mathcal{X}^{+}\right)^{2}}
            \partial\Sigma^{\prime}
\nonumber \\
 &  & \hphantom{\oint_{\tilde{\mathbf{z}}_{I}}
                 \frac{d\mathbf{z}}{2\pi i}
                 D\Phi\qquad}
   \left.\left.
   {}-\frac{\alpha^{2}}{2}
      \frac{D\mathcal{X}^{+}}{\partial^{2}\mathcal{X}^{+}}
      \partial\Sigma^{\prime} D\Sigma^{\prime}
    \vphantom{\frac{\partial^{3}D\mathcal{X}^{+}D\mathcal{X}^{+}}
                   {\left(\partial^{2}\mathcal{X}^{+}\right)^{2}}
              }
    \right]
    \frac{e^{-\alpha\Sigma^{\prime}}}
         {\left(\partial^{2}\mathcal{X}^{+}\right)^{\frac{\alpha}{4}}}
     \right|^{2}\ .
\end{eqnarray}
Replacing $\mathcal{X}^{+}$ by its expectation value, one can see
that $\mathcal{O}_{I}$ is equivalent to the insertion of 
\begin{equation}
\left| \frac{ \left[1-\alpha\frac{D\rho}{\partial^{2}\rho}
                       \partial D\Sigma^{\prime}
                    -\alpha\frac{\partial^{2}D\rho D\rho}
                                {\left(\partial^{2}\rho\right)^{2}}
                     \partial\Sigma^{\prime}
               \right]
               e^{-\alpha\Sigma^{\prime}}}
            { \left( \partial^{2}\rho
                     -\frac{5}{3}\frac{\partial^{3}D\rho D\rho}
                                      {\partial^{2}\rho}
                    +3\frac{\partial^{3}\rho\partial^{2}D\rho D\rho}
                           {\left(\partial^{2}\rho\right)^{2}}
              \right)^{\frac{\alpha}{4}}}
       \left(\tilde{\mathbf{z}}_{I}\right)
\right|^{2}\ .
\end{equation}

The ghost operator that should be inserted at $z=\infty$ can be seen
to be $e^{3\sigma^{\prime}-2\phi^{\prime}}$, whose superspace form
is 
\begin{equation}
\left(1+\theta\gamma b\right)
e^{3\sigma^{\prime}-2\phi^{\prime}}\left(z\right)\ .
\end{equation}
The conformal invariant combination which implements such an insertion
is given as 
\begin{equation}
\mathcal{R}
 \equiv
  \left| \oint_{\infty}\frac{d\mathbf{z}}{2\pi i}
           D\Phi
           \left(D\Theta^{+}\right)^{2\alpha}
           \left(1+\theta\gamma b\right)
           e^{3\sigma^{\prime}-2\phi^{\prime}}\left(z\right)
 \right|^{2}\ .
\end{equation}
At $\mathbf{z}=\mathbf{Z}_{r}$, one can see that 
$e^{\alpha\Sigma^{\prime}}\left(\mathbf{Z}_{r}\right)$
should be inserted.

The logarithmic singularities for $\mathcal{X}^{-}$ can be taken
care of by inserting 
\begin{equation}
\exp\left(\frac{d-10}{16}\frac{i}{p_{r}^{+}}\mathcal{X}^{+}\right)
    \left(\mathbf{Z}_{r,}\bar{\mathbf{Z}}_{r}\right)~,
\qquad
\exp\left(-\frac{d-10}{16}\frac{i}{p_{r}^{+}}\mathcal{X}^{+}\right)
   \left(\tilde{\mathbf{z}}_{I^{(r)}},
          \bar{\tilde{\mathbf{z}}}_{I^{(r)}}\right)\ .
\end{equation}
The latter should be made into the conformal invariant combination
\begin{equation}
\mathcal{S}_{r}
  \equiv \oint_{\tilde{\mathbf{z}}_{I^{\left(r\right)}}}
            \frac{d\mathbf{z}}{2\pi i}
            D\Phi
         \oint_{\bar{\tilde{\mathbf{z}}}_{I^{\left(r\right)}}}
            \frac{d\bar{\mathbf{z}}}{2\pi i}
            \bar{D}\tilde{\Phi}
          \exp\left(-\frac{d-10}{16}
                     \frac{i}{p_{r}^{+}}\mathcal{X}^{+}\right)
                      \left(\mathbf{z},\bar{\mathbf{z}}\right)\ .
\end{equation}

\subsection{Correlation functions in terms of the free variables}

The correlation functions can be expressed by using the free variables
with the insertions of the operators defined above. One can show the
supersymmetric version of eq.(\ref{eq:general}): 
\begin{eqnarray}
 &  & 
  \left\langle 
    \left|e^{3\sigma-2\phi}\left(\infty\right)\right|^{2}
    \phi_{1}\left(\mathbf{z}_{1},\bar{\mathbf{z}}_{1}\right)
    \phi_{2}\left(\mathbf{z}_{2},\bar{\mathbf{z}}_{2}\right)
     \cdots
    \phi_{n}\left(\mathbf{z}_{n},\bar{\mathbf{z}}_{n}\right)
    \prod_{r=1}^{N}
             e^{-ip_{r}^{+}\mathcal{X}^{-}}
             \left(\mathbf{Z}_{r},\bar{\mathbf{Z}}_{r}\right)
  \right\rangle _{\mathcal{X}^{\pm},B,C}
\nonumber \\
 &  & \quad
   =\mathcal{C} \left\langle \mathcal{R}
          \phi_{1}\left(\mathbf{z}_{1}\bar{\mathbf{z}}_{1}\right)
          \phi_{2}\left(\mathbf{z}_{2},\bar{\mathbf{z}}_{2}\right)
           \cdots
          \phi_{n}\left(\mathbf{z}_{n},\bar{\mathbf{z}}_{n}\right)
          \prod_{I=1}^{N-2}\mathcal{O}_{I}
       \right.
\nonumber \\
 &  & \quad
      \hphantom{\propto\quad}
   \left. \times
    \prod_{r=1}^{N}
        \left[ \mathcal{S}_{r}  \left|\alpha_{r}\right|^{-\alpha}
               \left|e^{\alpha\Sigma^{\prime}}(\mathbf{Z}_{r})
                \right|^{2}
               e^{-ip_{r}^{+}\mathcal{X}^{\prime-}
                   +\frac{d-10}{16}\frac{i}{p_{r}^{+}}\mathcal{X}^{+}}
                       \left(\mathbf{Z}_{r},\bar{\mathbf{Z}}_{r}\right)
         \right]
     \right\rangle _{\mathrm{free}}.
\label{eq:supergeneral}
\end{eqnarray}
Here $\phi_{i}\ \left(i=1,\cdots,n\right)$ are made of 
$\mathcal{X}^{+}$, $D\mathcal{X}^{-}$, $\bar{D}\mathcal{X}^{-}$,
$B$, $C$, $\tilde{B}$, $\tilde{C}$
and their covariant derivatives
and $\mathcal{C}$ is a numerical constant.

This formula can be shown in the same way as eq.(\ref{eq:general})
in the bosonic case. One can first prove the simplest case 
\begin{eqnarray}
\lefteqn{
   \left\langle 
     \left|e^{3\sigma-2\phi}\left(\infty\right)\right|^{2}
     \prod_{r=1}^{N} e^{-ip_{r}^{+}\mathcal{X}^{-}}
                      \left(\mathbf{Z}_{r},\bar{\mathbf{Z}}_{r}\right)
   \right\rangle _{\mathcal{X}^{\pm},B,C}
}  \nonumber \\
&  & 
  \propto
  \left\langle \mathcal{R} \prod_{I=1}^{N-2}\mathcal{O}_{I}
       \prod_{r=1}^{N}
           \left[ \mathcal{S}_{r}
                  \left|\alpha_{r}\right|^{-\alpha}
                  \left|e^{\alpha\Sigma^{\prime}}(\mathbf{Z}_{r})
                    \right|^{2}
                  e^{-ip_{r}^{+}\mathcal{X}^{\prime-}
                     +\frac{d-10}{16}\frac{i}{p_{r}^{+}}\mathcal{X}^{+}}
                       \left(\mathbf{Z}_{r},\bar{\mathbf{Z}}_{r}\right)
            \right]
   \right\rangle _{\mathrm{free}}.~~
\label{eq:supercorrelationfree}
\end{eqnarray}
The left hand side was evaluated in Ref.~\cite{Baba:2009fi} to
be $e^{-\alpha\Gamma_{\mathrm{super}}}$ and the right hand side is
easily computed to be proportional to it. It is much more complicated
but straightforward to prove the supersymmetric versions of 
eqs.(\ref{eq:X-0})
and (\ref{eq:delX-}) and show that eq.(\ref{eq:supergeneral}) holds.
From eq.(\ref{eq:supergeneral}), one can see that 
$\mathcal{R},\mathcal{O}_I,\mathcal{S}_r$ given in the previous subsection 
are conformal invariant. 

Using eq.(\ref{eq:supergeneral}), one can rewrite arbitrary correlation
functions in the $X^{\pm}$ CFT using the free theory and vice versa.
In particular, one can modify the vertex operator at 
$\mathbf{z}=\mathbf{Z}_{r}$,
by taking the limit $\mathbf{z}_{i}\to\mathbf{Z}_{r}$ appropriately.
Thus we can get various vertex operators in the $X^{\pm}$ CFT and
their free field versions. Hence, to any vertex operator in the
free field description there exists a corresponding vertex operator
in the $X^{\pm}$ CFT, and vice versa.


\section{Vertex operators}
\label{sec:vo}

In the conformal gauge, 
the light-cone gauge noncritical superstrings can be described by
worldsheet theory which consists of the supersymmetric $X^{\pm}$
CFT, super-reparametrization ghosts and the transverse variables.
We have shown that the former two systems combined can be described 
by free variables in the previous section. 
Vertex operators are made from these variables and should be BRST
invariant. We would like to construct BRST invariant vertex operators
in the Ramond sector using the free variables.
Before doing so, we examine how the BRST invariant vertex operators
in the Neveu-Schwarz sector are expressed in terms of the free variables.
We construct the BRST invariant vertex operators in the Ramond sector 
imitating those in the Neveu-Schwarz sector.

\subsection{Vertex operators in the Neveu-Schwarz sector}

Let us consider the left-moving part of a state in the Neveu-Schwarz
sector of the light-cone gauge superstrings:
\begin{equation}
\alpha_{-n_{1}}^{i_{1}}\cdots\psi_{-s_{1}}^{j_{1}}\cdots
  \left|\vec{p}\right\rangle _{L}\ .
\label{eq:LCstate-NS}
\end{equation}
Here $n_{i}$ are positive integers and $s_{i}$ are positive half
odd integers. $|\vec{p}\rangle_{L}$ is the state which
corresponds to the
operator $e^{i\vec{p}\cdot\vec{X}_{L}}$, where $\vec{p}=(p^{i})$
$(i=1,\ldots,d-2)$ denotes the transverse $(d-2)$-momentum 
and $\vec{X}_{L}$
denotes the left-moving part of the transverse variables 
$\vec{X}=(X^{i})$.

The left-moving BRST invariant vertex operator in the conformal gauge
corresponding to this state is given as\ \cite{Baba:2009zm}
\begin{eqnarray}
V_{L}^{\left(-1\right)} (z)
& \equiv & A_{-n_{1}}^{i_{1}}\cdots B_{-s_{1}}^{j_{1}}
          \cdots
\nonumber \\
 &  & \ \times 
   e^{\sigma-\phi}
   \exp\left[-ip^{+}X_{L}^{-}
             -i\left(p^{-}-\frac{\mathcal{N}}{p^{+}}
                     +\frac{d-10}{16}\frac{1}{p^{+}}
               \right)X_{L}^{+}
             +i\vec{p}\cdot\vec{X}_{L}
        \right]  \left(z\right),~~~
\label{eq:VL-1}
\end{eqnarray}
where $A_{-n}^{i}$ and $B_{-s}^{i}$ are the DDF operators 
defined as
\begin{eqnarray}
A_{-n}^{i} 
& \equiv & 
  \oint_{z}\frac{dz^{\prime}}{2\pi i}
      \left(i\partial X^{i}+\frac{n}{p^{+}}\psi^{i}\psi^{+}\right)
      e^{-i\frac{n}{p^{+}}X_{L}^{+}}
      \left(z^{\prime}\right)\ ,
\nonumber \\
B_{-s}^{i} 
& \equiv & 
  \oint_{z}\frac{dz^{\prime}}{2\pi i}
   \left(\psi^{i} -\partial X^{i}\frac{\psi^{+}}{\partial X^{+}}
         -\frac{1}{2}\psi^{i}
            \frac{\psi^{+}\partial\psi^{+}}
                 {\left(\partial X^{+}\right)^{2}}
   \right)
   \left( \frac{i\partial X^{+}}{p^{+}}\right)^{\frac{1}{2}}
   e^{-i\frac{s}{p^{+}}X_{L}^{+}}\left(z^{\prime}\right)\ ,
\label{eq:DDF}
\end{eqnarray}
$\mathcal{N}\equiv\sum_{i}n_{i}+\sum_{j}s_{j}$, and $p^{-}$ is
taken to satisfy the on-shell condition
\begin{equation}
p^{-}  
 =  \frac{1}{p^{+}}
    \left(\frac{1}{2}\vec{p}^{2}+\mathcal{N}-\frac{d-2}{16}\right)\ .
\label{eq:-1onshell}
\end{equation}
The superscript $\left(-1\right)$ on the left hand side of 
eq.(\ref{eq:VL-1}) indicates the picture number. 
$V_{L}^{\left(-1\right)}$ appears to
have the momentum in the $-$ direction shifted as 
$p^{-}+\frac{d-10}{16}\frac{1}{p^{+}}$
instead of $p^{-}$. 
Because of the shift, the vertex operator $V_{L}^{\left(-1\right)}$
becomes of dimension $0$ and BRST invariant. 
In the scattering amplitudes,
this shift comes with the insertion of 
$\exp\left(i\frac{d-10}{16}\frac{1}{p^{+}}X^{+}\right)$
at the interaction points\ \cite{Baba:2009ns}, 
and thus the momentum conserved is $p^{-}$.

The free field expression 
$V_{L}^{\prime\left(-1\right)}$ for $V_{L}^{\left(-1\right)}$
can be read off from eq.(\ref{eq:supergeneral}) and 
\begin{eqnarray}
V_{L}^{\prime\left(-1\right)} (z)
& = & A_{-n_{1}}^{i_{1}}\cdots B_{-s_{1}}^{j_{1}} \cdots
\nonumber \\
 &  & \ \times e^{\left(1+\alpha\right)
                   \left(\sigma^{\prime}-\phi^{\prime}\right)}
   \exp\left[ -ip^{+}X_{L}^{\prime-}
              -i\left( p^{-}-\frac{\mathcal{N}}{p^{+}}
                \right)  X_{L}^{+}
              +i\vec{p} \cdot \vec{X}_{L}
       \right]  \left(z\right),
\label{eq:Vprime-1}
\end{eqnarray}
up to a factor which depends on $\alpha_{r}$. 
Since the variable $\mathcal{X}^{+}$ is common between 
the $X^{\pm}$ CFT and the free theory, 
the DDF operators $A_{-n}^{i}$ and $B_{-s}^{i}$ can be defined
in the same way for the both theories. 
One good feature of $V_{L}^{\prime\left(-1\right)}$
is that unlike $V_{L}^{(-1)}$ the momentum $p^{-}$ is not shifted.
With the ghost factor 
$e^{\left(1+\alpha\right)
      \left(\sigma^{\prime}-\phi^{\prime}\right)}$,
one can see that the dimension of $V_{L}^{\prime\left(-1\right)}$
is $0$ when the on-shell condition~(\ref{eq:-1onshell}) 
is satisfied.
The insertions of $\mathcal{S}_{r}$ in the free field description
take care of the momentum conservation.

The right-moving vertex operator $V_{R}^{\prime(-1)}$ can be 
constructed in the same way.

\subsection{Vertex operators in the Ramond sector \label{sec:vo-ramond}}

Using the free variables, it is easy to construct the spin fields.
We can bosonize $\psi^{+}$ and $\psi^{\prime-}$ as 
\begin{equation}
\psi^{+}(z)=e^{iH^{\prime}}(z)\ ,
\qquad
\psi^{\prime-}(z)=-e^{-iH^{\prime}}(z)\ ,
\label{eq:psiprime-bosonize}
\end{equation}
with 
$H^{\prime}\left(z\right)H^{\prime}\left(w\right)
   \sim-\ln\left(z-w\right)$.
This yields 
\begin{equation}
\psi^{\prime-}\psi^{+}(z)=i\partial H^{\prime}(z)~.
\end{equation}
Using $H^{\prime}$, we can construct the spin field 
for the longitudinal variables as 
\begin{equation}
e^{\pm\frac{i}{2}H^{\prime}}\ .
\end{equation}
The ghost variables $b'$, $c'$, $\beta'$, $\gamma'$ are bosonized
as in eq.(\ref{eq:betaprime-bosonize}). 
The ghost part of the Ramond vertex operators 
is given by the conformal primary field 
\begin{equation}
e^{\left(1+\alpha\right)
      \left(\sigma^{\prime}-\phi^{\prime}\right)
   \pm \frac{1}{2}\phi^{\prime}}
\label{eq:vo-ghost-Ramond}
\end{equation}
with weight $-\frac{1}{2}\alpha-\frac{5}{8}$.

Eq.(\ref{eq:supergeneral}) holds even when some of $\phi_{i}$ involve
such spin fields. For example, let us consider a pair of spin fields
$e^{\frac{i}{2}H^{\prime}}\left(z\right)
  e^{-\frac{i}{2}H^{\prime}}\left(w\right)$.
It can be expressed as 
\begin{equation}
e^{\frac{i}{2}H^{\prime}}\left(z\right)
 e^{-\frac{i}{2}H^{\prime}}\left(w\right)
=(z-w)^{-\frac{1}{4}}
 \sum_{n=0}^{\infty}\frac{1}{n!}
    :\left(\frac{1}{2}\int_{w}^{z}dz^{\prime}\,
            i\partial H^{\prime}\left(z^{\prime}\right)
     \right)^{n}:\ ,
\label{eq:spincorrespondence}
\end{equation}
in terms of $i\partial H^{\prime}$, to which eq.(\ref{eq:supergeneral})
is applicable. Since spin fields always appear in such pairs in the
correlation function, one can see that eq.(\ref{eq:supergeneral})
holds even in the presence of the spin fields. Taking the limit mentioned
at the end of the last section, we can get the vertex operators
containing the spin fields.

\subsubsection*{BRST invariant vertex operator}

%
%
%

With these spin fields, we can construct BRST invariant vertex operators
in the Ramond sector.
Let us consider the left-moving part of a state in the Ramond sector
of the light-cone gauge superstring of the form 
\begin{equation}
\alpha_{-n_{1}}^{i_{1}}\cdots\psi_{-m_{1}}^{j_{1}}\cdots
   \left|\vec{p},\vec{s}\right\rangle _{L}\ ,
\label{eq:LCstate-Ramond}
\end{equation}
where $n_{i}$ and $m_{i}$ are positive integers, and 
$\left|\vec{p},\vec{s}\right\rangle _{L}$
is the state corresponding to the operator 
$e^{i\vec{p}\cdot\vec{X}_{L}+i\vec{s}\cdot\vec{H}}$.
Here $\vec{H}=(H^{A})$ $\left(A=1,\ldots,\frac{d-2}{2}\right)$ are
defined by using the transverse fermions as 
\begin{equation}
e^{\pm iH^{A}}
 =\frac{1}{\sqrt{2}}\left(\psi^{2A-1}\pm i\psi^{2A}\right)~,
\end{equation}
and $\vec{s}=(s^{A})$ with $s^{A}=\frac{1}{2}$ or $-\frac{1}{2}$.

In order to express the vertex operator for the Ramond sector state
(\ref{eq:LCstate-Ramond}), we need to use the free fields. Imitating
$V_{L}^{\prime\left(-1\right)}$ in eq.(\ref{eq:Vprime-1}), we construct
\begin{eqnarray}
V_{L}^{\prime\left(-\frac{3}{2}\right)}\left(z\right) 
& \equiv & 
  A_{-n_{1}}^{i_{1}}\cdots B_{-m_{1}}^{j_{1}} \cdots
  \exp\left[\frac{i}{2}H^{\prime}+i\vec{s}\cdot\vec{H}\right]
      \left(z\right)
\nonumber \\
 &  & \ \times 
   e^{\left(1+\alpha\right)\left(\sigma^{\prime}-\phi^{\prime}\right)
      -\frac{1}{2}\phi^{\prime}}
   \exp\left[ -ip^{+}X_{L}^{\prime-}
              -i\left(p^{-}-\frac{\mathcal{N}}{p^{+}}\right)X_{L}^{+}
              +i\vec{p}\cdot\vec{X}_{L}
       \right]  \left(z\right),
\nonumber \\
V_{L}^{\prime\prime\left(-\frac{1}{2}\right)}\left(z\right) 
& \equiv & A_{-n_{1}}^{i_{1}} \cdots B_{-m_{1}}^{j_{1}} \cdots
  \exp\left[ -\frac{i}{2}H^{\prime}+i\vec{s}\cdot\vec{H}\right]
        \left(z\right)
\nonumber \\
 &  & \ \times 
   e^{\left(1+\alpha\right)\left(\sigma^{\prime}-\phi^{\prime}\right)
      +\frac{1}{2}\phi^{\prime}}
   \exp\left[ -ip^{+}X_{L}^{\prime-}
              -i\left(p^{-}-\frac{\mathcal{N}}{p^{+}}\right)X_{L}^{+}
             +i\vec{p}\cdot\vec{X}_{L}
        \right]   \left(z\right),
\nonumber \\
\label{eq:VDDF-L-prime}
\end{eqnarray}
where $\mathcal{N}\equiv\sum_{i}n_{i}+\sum_{j}m_{j}$, 
and $A_{-n}^{i}$ and $B_{-m}^{i}$ are the DDF operators 
defined in eq.(\ref{eq:DDF})
with $s$ replaced by an integer $m$. 
The reason why we adopt the notation 
$V_{L}^{\prime\prime\left(-\frac{1}{2}\right)}\left(z\right)$ 
for the second one will become clear later. 
Since the conformal dimension
of the transverse spin field $e^{i\vec{s}\cdot\vec{H}}$ is 
$\frac{d-2}{16}$,
this time the on-shell condition is 
\begin{equation}
p^{-}
=\frac{1}{p^{+}}\left(\frac{1}{2}\vec{p}^{2}+\mathcal{N}\right)\ .
\label{eq:onshell-1-2}
\end{equation}
 One can construct the right-moving counterparts 
$V_{R}^{\prime\left(-\frac{3}{2}\right)}\left(\bar{z}\right)$
and 
$V_{R}^{\prime\prime\left(-\frac{1}{2}\right)}\left(\bar{z}\right)$
in the same way.

As we argued above, there should exist a vertex operator 
$V_{L}^{\left(-\frac{3}{2}\right)}$
in the $X^{\pm}$ CFT which corresponds to 
$V_{L}^{\prime\left(-\frac{3}{2}\right)}$
in the free field description. The explicit form of 
$V_{L}^{\left(-\frac{3}{2}\right)}$
will be complicated because of the presence of the spin fields, but
even without the explicit form, we can read off its properties from
its free field form $V_{L}^{\prime\left(-\frac{3}{2}\right)}$. 
In particular, one
can show that $V_{L}^{\left(-\frac{3}{2}\right)}$ is BRST invariant.
Using the explicit form of $D\Theta^{+}$ in terms of the component
fields, 
\begin{equation}
D\Theta^{+}
  =\left(\partial X^{+}\right)^{\frac{1}{2}}
   \left[ 1+\frac{\partial\psi^{+}\psi^{+}}
                 {2\left(\partial X^{+}\right)^{2}}
            +\theta
              \left(\frac{i\partial\psi^{+}}{\partial X^{+}}
                     -\frac{i}{2}
                       \frac{\partial^{2}X^{+}}
                             {\left(\partial X^{+}\right)^{2}}
                       \psi^{+}
               \right)
    \right]\ ,
\end{equation}
one can obtain 
\begin{equation}
D\Theta^{+}\left(z\right)
  V_{L}^{\prime\left(-\frac{3}{2}\right)}\left(0\right)
\sim
  \left(\frac{-ip^{+}}{z}\right)^{\frac{1}{2}}
   V_{L}^{\prime\left(-\frac{3}{2}\right)}\left(0\right)\ .
\end{equation}
Combining this relation with eqs.(\ref{eq:ghost-sf}), 
(\ref{eq:Xprime-}),
and the OPE's between the primed ghost fields and 
$V_{r}^{\prime\left(-\frac{3}{2}\right)}$,
we have 
\begin{eqnarray}
b\left(z\right)V_{L}^{\left(-\frac{3}{2}\right)}\left(0\right) 
  &\sim& z^{-1}b_{-1}V_{L}^{\left(-\frac{3}{2}\right)}\left(0\right)\ ,
 \nonumber \\
c\left(z\right)V_{L}^{\left(-\frac{3}{2}\right)}\left(0\right) 
  &\sim& z c_{0}V_{L}^{\left(-\frac{3}{2}\right)}\left(0\right)\ ,
\nonumber \\
\beta\left(z\right)V_{L}^{\left(-\frac{3}{2}\right)}\left(0\right) 
  &\sim& z^{-\frac{3}{2}}\beta_{0}
         V_{L}^{\left(-\frac{3}{2}\right)}\left(0\right)\ ,
\nonumber \\
\gamma\left(z\right)V_{L}^{\left(-\frac{3}{2}\right)}\left(0\right) 
  &\sim& z^{\frac{3}{2}}\gamma_{-1}
         V_{L}^{\left(-\frac{3}{2}\right)}\left(0\right)\ .
\end{eqnarray}
These relations imply that $V_{L}^{\left(-\frac{3}{2}\right)}$ should
be of the form 
\begin{equation}
e^{\sigma-\frac{3}{2}\phi}\mathcal{O}_{X}\ ,
\label{eq:OX}
\end{equation}
where $\mathcal{O}_{X}$ is a conformal field made of the unprimed
longitudinal and the transverse variables, which satisfies 
\begin{eqnarray}
T_{B}^{X}(z)\mathcal{O}_{X}\left(0\right)
   & \sim & 
      \frac{\frac{1}{2}p^{2}+\mathcal{N}+\frac{5}{8}}{z^{2}}
      \mathcal{O}_{X}(0)
      + \frac{1}{z}\partial\mathcal{O}_{X}(0)\ ,
\nonumber \\
T_{F}^{X}(z)\mathcal{O}_{X}\left(0\right) 
    & \sim & z^{-\frac{3}{2}}G_0^X\mathcal{O}_X(0)\ ,
\label{eq:OPE-TB-TF}
\end{eqnarray}
where $T_{B}^{X},T_{F}^{X},G_0^X$ are the energy-momentum tensor, 
the supercurrent and the supercharge in the matter respectively. 
Using eqs.(\ref{eq:OX})
and (\ref{eq:OPE-TB-TF}), we can prove that 
$V^{\left(-\frac{3}{2}\right)}$
commutes with the BRST operator 
\begin{equation}
Q_{\mathrm{B}}
  =\oint\frac{dz}{2\pi i}
     \left[cT_{B}^{X} -\gamma T_{F}^{X}
           -\frac{1}{2}c\gamma\partial\beta
           -\frac{3}{2}c\partial\gamma\beta
           -bc\partial c
           -\frac{1}{4}b\gamma^{2}\right]
+\mathrm{c.c.}\ ,
\end{equation}
if the on-shell condition (\ref{eq:onshell-1-2}) is satisfied.

Unfortunately, $V_{L}^{\prime\prime\left(-\frac{1}{2}\right)}$ does
not correspond to a BRST invariant vertex operator. 
We can obtain
the left-moving BRST invariant vertex operator 
$V_{L}^{\left(-\frac{1}{2}\right)}$
in the $-\frac{1}{2}$ picture by applying to 
$V_{L}^{\left(-\frac{3}{2}\right)}$
the picture changing operator $X(z)$ defined as 
\begin{equation}
X\left(z\right)
   \equiv \left\{ Q_{\mathrm{B}}\,,\,\xi\left(z\right)\right\} 
   = c\partial\xi
      -e^{\phi}T_{F}^{X}
      +\frac{1}{4}\partial b\eta e^{2\phi}
      +\frac{1}{4}b\left(2\partial\eta e^{2\phi}
      +\eta\partial e^{2\phi}\right)\ ,
\end{equation}
namely 
\begin{equation}
V_{L}^{\left(-\frac{1}{2}\right)}\left(0\right)
   \equiv  \lim_{z\to0} X(z)
               V_{L}^{\left(-\frac{3}{2}\right)} \left(0\right)
   = \lim_{z\to0} \left(-e^{\phi}T_{F}^{X}\right)(z)
        V_{L}^{\left(-\frac{3}{2}\right)}\left(0\right)\ .
\label{eq:VLr12}
\end{equation}
We define $V_{L}^{\prime\left(-\frac{1}{2}\right)}$ to be the free
field version of $V_{L}^{\left(-\frac{1}{2}\right)}$. 

One can obtain the BRST invariant right-moving parts 
$V_{R}^{\left(-\frac{3}{2}\right)}\left(\bar{z}\right)$,
$V_{R}^{\left(-\frac{1}{2}\right)}\left(\bar{z}\right)$
and their free field versions in the same way.


\section{Amplitudes\label{sec:Amplitudes}}

In this section, we would like to show that the tree level amplitudes
in the noncritical light-cone gauge string field theory can be expressed
by using the BRST invariant vertex operators constructed in the previous
section. 
Our procedure is as follows. 
We start  from the light-cone gauge amplitudes.
We rewrite them by adding the longitudinal 
and ghost degrees of freedom.
Then we reach the BRST invariant conformal gauge expression.

\subsection{Light-cone gauge superstring field theory in $d$ dimensions}

The light-cone gauge superstring field theory can be defined in noncritical
dimensions by just considering the action with three-string interactions
like that for the (NS,NS) strings in Ref.~\cite{Baba:2009zm}. However,
putting naively $d\ne10$ makes the physical content of the theory
quite different from that in the critical case. 
{}From the on-shell conditions~(\ref{eq:-1onshell}) 
and (\ref{eq:onshell-1-2})
for the Neveu-Schwarz and the Ramond sectors, 
the level matching condition for the (NS,R) sector becomes
\begin{equation}
\mathcal{N} =  \tilde{\mathcal{N}}+\frac{d-2}{16}\ ,
\end{equation}
where $\mathcal{N}$ and $\tilde{\mathcal{N}}$ denote the level
numbers in the left- and the right-moving parts. 
Since $\mathcal{N}-\tilde{\mathcal{N}}$
is a half-integer, there exist no states satisfying this condition
for generic $d$. The situation is the same for the (R,NS) sector
and one can see that the theory does not include any spacetime fermions
for generic $d$. This fact is problematic if one wants to use the
noncritical string theory to dimensionally regularize the critical
theory. We need to modify the worldsheet theory for such applications.
We will deal with this problem elsewhere. Here we take the theory
as it is and consider the theory for generic $d\ne10$ with only (NS,NS)
and (R,R) sectors%
\footnote{We might have to consider Type~$0$ theory for $d\ne10$ 
          for the modular invariance 
          but as far as we are discussing tree amplitudes, 
          there is not so big difference between Type~II theory and 
          Type~$0$ theory.}
and calculate the tree amplitudes. 
In appendix~\ref{sec:SFTaction},
we present the string field theory action in this situation.

The tree level $N$-string amplitudes $\mathcal{A}_{N}$ are perturbatively
computed in the same way as those in Ref.~\cite{Baba:2009zm}. 
Starting from the action (\ref{eq:sftaction}) of string field theory, 
we obtain
\begin{equation}
\mathcal{A}_{N}
 =\left(4ig\right)^{N-2}
  \int\left(\prod_{\mathcal{I}=1}^{N-3}
              \frac{d^{2}\mathcal{T}_{\mathcal{I}}}{4\pi}\right)
   F_{N}\left(\mathcal{T}_{\mathcal{I}},
              \bar{\mathcal{T}}_{\mathcal{I}}\right),
\label{eq:AN}
\end{equation}
where $\mathcal{T}_{\mathcal{I}}$ denotes the complex Schwinger
parameter of the $\mathcal{I}$th internal propagator 
$(\mathcal{I}=1,\ldots,N-3)$,
which consists of the $N-3$ complex moduli parameters of the amplitude
$\mathcal{A}_{N}$. As was discussed in Ref.~\cite{Baba:2009zm},
on the right hand side the integration region is taken to cover the
whole moduli space and the integrand $F_{N}$ is described by the
correlation function of the superconformal field theory for the light-cone
gauge superstrings on the $z$-plane: 
\begin{eqnarray}
F_{N}\left(\mathcal{T}_{\mathcal{I}},
           \bar{\mathcal{T}}_{\mathcal{I}}\right)
 & = & \left(2\pi\right)^{2}
       \delta\left(\sum_{r=1}^{N}p_{r}^{+}\right)
       \delta\left(\sum_{r=1}^{N}p_{r}^{-}\right)
       \mathrm{sgn}\left(\prod_{r=1}^{N}\alpha_{r}\right)
       e^{-\frac{d-2}{16}\Gamma}\nonumber \\
 &  & \qquad\times
    \left\langle \prod_{I=1}^{N-2}
                   \left|\left(\partial^{2}\rho\right)^{-\frac{3}{4}}
                         T_{F}^{\mathrm{LC}}\left(z_{I}\right)
                   \right|^{2}
                \prod_{r=1}^{N}V_{r}^{\mathrm{LC}}
    \right\rangle _{X^{i}}~.
\label{eq:FN}
\end{eqnarray}
Here $\Gamma$ is given in eq.(\ref{eq:Gamma}), and 
$V_{r}^{\mathrm{LC}}$
denotes the vertex operators for the $r$th external string in the
light-cone gauge. An external state in the (NS,NS) sector is obtained
by multiplying the state (\ref{eq:LCstate-NS}) by a similar one in
the right-moving sector: 
\begin{equation}
\alpha_{-n_{1}}^{i_{1}\left(r\right)}
   \cdots
\tilde{\alpha}_{-\tilde{n}_{1}}^{\tilde{\imath}_{1}\left(r\right)}
   \cdots
\psi_{-s_{1}}^{j_{1}\left(r\right)}
   \cdots
\tilde{\psi}_{-\tilde{s}_{1}}^{\tilde{\jmath}_{1}\left(r\right)}
  \cdots
\left|\vec{p_{r}}\right\rangle _{r}\ .
\end{equation}
To this state corresponds a vertex operator 
\begin{eqnarray}
V_{r}^{\mathrm{LC}} 
  & = & \alpha_{r}
    \oint_{0}\frac{dw_{r}}{2\pi i}
        \partial_{w_{r}}X^{i_{1}}\left(w_{r}\right)w_{r}^{-n_{1}}
      \cdots
    \oint_{0}\frac{d\bar{w}_{r}}{2\pi i}
         \partial_{\bar{w}_{r}}X^{\tilde{\imath}_{1}}
              \left(\bar{w}_{r}\right)\bar{w}_{r}^{-\tilde{n}_{1}}
        \cdots
\nonumber \\
 &  & \quad\times
   \oint_{0}\frac{dw_{r}}{2\pi i}
         \psi^{j_{1}}\left(w_{r}\right)w_{r}^{-s_{1}-\frac{1}{2}}
       \cdots
   \oint_{0}\frac{d\bar{w}_{r}}{2\pi i}
         \tilde{\psi}^{\tilde{\jmath}_{1}}\left(\bar{w}_{r}\right)
          \bar{w}_{r}^{-\tilde{s}_{1}-\frac{1}{2}}
        \cdots
\nonumber \\
 &  & \quad\times 
    e^{i\vec{p}_{r}\cdot\vec{X}}\left(w_{r}=0,\bar{w}_{r}=0\right)
    e^{-p_{r}^{-}\tau_{0}^{\left(r\right)}}\ ,
\end{eqnarray}
where $w_{r}$ is the local coordinate, introduced in the region
$z\sim Z_{r}$ as 
\begin{equation}
w_{r} \equiv
  \exp\left[ \frac{1}{\alpha_{r}}
              \left(\rho-\tau_{0}^{(r)}-i\beta_{r} \right)
       \right]~,
\qquad
 \tau_{0}^{(r)}+i\beta_{r}\equiv\rho(z_{I^{(r)}})~.
\end{equation}
Similarly, an (R,R) external state is obtained by multiplying the
state~(\ref{eq:LCstate-Ramond}) by a similar one in the right-moving
sector: \begin{equation}
\alpha_{-n_{1}}^{i_{1}\left(r\right)}
  \cdots
\tilde{\alpha}_{-\tilde{n}_{1}}^{\tilde{\imath}_{1}\left(r\right)}
  \cdots
\psi_{-m_{1}}^{j_{1}\left(r\right)}
  \cdots
\tilde{\psi}_{-\tilde{m}_{1}}^{\tilde{\jmath}_{1}\left(r\right)}
  \cdots
\left|\vec{p}_{r},\vec{s}_{r},\vec{\tilde{s}}_{r}\right\rangle _{r}\ .
\end{equation}
For this state, we should take 
\begin{eqnarray}
V_{r}^{\mathrm{LC}} 
  & = & \alpha_{r}
     \oint_{0}\frac{dw_{r}}{2\pi i}
         \partial_{w_{r}}X^{i_{1}}\left(w_{r}\right)w_{r}^{-n_{1}}
             \cdots
     \oint_{0}\frac{d\bar{w}_{r}}{2\pi i}
         \partial_{\bar{w}_{r}}X^{\tilde{\imath}_{1}}
              \left(\bar{w}_{r}\right)
         \bar{w}_{r}^{-\tilde{n}_{1}}
               \cdots
\nonumber \\
 &  & \quad\times
    \oint_{0}\frac{dw_{r}}{2\pi i}\psi^{j_{1}}\left(w_{r}\right)
                 w_{r}^{-m_{1}-\frac{1}{2}}
          \cdots
     \oint_{0}\frac{d\bar{w}_{r}}{2\pi i}
          \tilde{\psi}^{\tilde{\jmath}_{1}}\left(\bar{w}_{r}\right)
                \bar{w}_{r}^{-\tilde{m}_{1}-\frac{1}{2}}
         \cdots
\nonumber \\
 &  & \quad\times 
    e^{i\vec{p}_{r}\cdot\vec{X}
       +i\vec{s}_{r}\cdot\vec{H}
       +i\vec{\tilde{s}}_{r}\cdot\vec{\tilde{H}}}
           \left(w_{r}=0,\bar{w}_{r}=0\right)
    e^{-p_{r}^{-}\tau_{0}^{\left(r\right)}}\ .
\end{eqnarray}

\subsection{Longitudinal variables and ghosts}

We rewrite the light-cone gauge expression (\ref{eq:AN}) by adding
the longitudinal variables and the super-reparametrization ghosts to the
worldsheet theory. 
Suppose that $V_{r}^{\mathrm{LC}}\ \left(r=1,\cdots,2f\right)$
are in the (R,R) sector and the other $V_{r}^{\mathrm{LC}}$'s are
in the (NS,NS) sector. It is straightforward to show that the quantity
which appears on the right hand side of eq.(\ref{eq:FN}) can be expressed
as a correlation function in the system of the free variables defined
in section \ref{sec:supersymmetric-case}:
\begin{eqnarray}
\lefteqn{
   (2\pi)^{2} \delta\left(\sum_{r=1}^{N}p_{r}^{+}\right)
              \delta\left(\sum_{r=1}^{N}p_{r}^{-}\right)
   e^{-\frac{d-2}{16}\Gamma}
   \prod_{I=1}^{N-2}
       \left|\partial^{2}\rho\left(z_{I}\right)\right|^{-\frac{3}{2}}
   \prod_{r=1}^{N} V_{r}^{\mathrm{LC}}
}  \nonumber \\
&  & \sim
   \left\langle 
      \left|(\partial\rho)^{1+\alpha}e^{\sigma^{\prime}}(\infty)
      \right|^{2}
      \prod_{I=1}^{N-2}
         \left|\frac{e^{-\left(1+\alpha\right)
                         \left(\sigma^{\prime}-\phi^{\prime}\right)}}
                    {\left(\partial^{2}\rho\right)^{1+\frac{\alpha}{4}}}
                 (z_{I})
          \right|^{2}
       \prod_{r=1}^{f}
           \left( |\alpha_{r}|^{-\alpha}
                  V_{r}^{\prime\left(-\frac{3}{2},-\frac{3}{2}\right)}
                      \left(Z_{r},\bar{Z}_{r}\right)
            \right)
     \right.
\nonumber \\
 &  & \hphantom{\sim\quad}
   \left.\times
   \prod_{r=f+1}^{2f}
         \left(|\alpha_{r}|^{-\alpha}
               V_{r}^{\prime\prime\left(-\frac{1}{2},-\frac{1}{2}\right)}
                   \left(Z_{r},\bar{Z}_{r}\right)
          \right)
    \prod_{r=2f+1}^{N}
          \left(|\alpha_{r}|^{-\alpha}
                V_{r}^{\prime\left(-1,-1\right)}
                    \left(Z_{r},\bar{Z}_{r}\right)
          \right)
  \right\rangle _{\mathrm{free}}.
\label{eq:VDDF-VLC-RNS}
\end{eqnarray}
Here 
\begin{eqnarray}
V_{r}^{\prime\left(-1,-1\right)}(Z_{r},\bar{Z}_{r})
 & \equiv & V_{L,r}^{\prime\left(-1\right)}(Z_{r})
            V_{R,r}^{\prime\left(-1\right)}(\bar{Z}_{r})~,
\nonumber \\
V_{r}^{\prime\left(-\frac{3}{2},-\frac{3}{2}\right)}(Z_{r},\bar{Z}_{r})
  & \equiv & |\alpha_{r}|^{-(\alpha+1)}
             V_{L,r}^{\prime\left(-\frac{3}{2}\right)}(Z_{r})
             V_{R,r}^{\prime\left(-\frac{3}{2}\right)}(\bar{Z}_{r})~,
\nonumber \\
V_{r}^{\prime\prime\left(-\frac{1}{2},-\frac{1}{2}\right)}
    (Z_{r},\bar{Z}_{r}) 
  & \equiv & |\alpha_{r}|^{\alpha+1}
             V_{L,r}^{\prime\prime\left(-\frac{1}{2}\right)}(Z_{r})
             V_{R,r}^{\prime\prime\left(-\frac{1}{2}\right)}
                   (\bar{Z}_{r})~,
\end{eqnarray}
where $V_{L,r}^{\prime\left(-1\right)}$, $V_{L,r}^{\prime\left(-\frac{3}{2}\right)}$
and $V_{L,r}^{\prime\prime\left(-\frac{1}{2}\right)}$ are the vertex
operators $V_{L}^{\prime\left(-1\right)}$, 
$V_{L}^{\prime\left(-\frac{3}{2}\right)}$
and $V_{L}^{\prime\prime\left(-\frac{1}{2}\right)}$ 
defined in eqs.(\ref{eq:Vprime-1})
and (\ref{eq:VDDF-L-prime}) for the $r$-th external string, and
similarly for the right moving sector ones 
$V_{R,r}^{\prime\left(-1\right)}$,
$V_{R,r}^{\prime\left(-\frac{3}{2}\right)}$ and 
$V_{R,r}^{\prime\prime\left(-\frac{1}{2}\right)}$.
In deriving eq.(\ref{eq:VDDF-VLC-RNS}), we have used the relation
\begin{equation}
\frac{\prod_{f+1\leq s<r\leq2f} \left|Z_{r}-Z_{s}\right|^{2}
      \cdot
      \prod_{r=1}^{f}\prod_{I} \left| Z_{r}-z_{I}\right|
      \cdot
      \prod_{r=f+1}^{2f}\prod_{s=2f+1}^{N}
          \left| Z_{r}-Z_{s}\right|}
      {\prod_{1\leq s<r\leq f} \left|Z_{r}-Z_{s}\right|^{2}
       \cdot
       \prod_{r=f+1}^{2f}\prod_{I} \left| Z_{r}-z_{I}\right|
      \cdot
       \prod_{r=1}^{f}\prod_{s=2f+1}^{N} \left|Z_{r}-Z_{s}\right|}
= \frac{\prod_{r=1}^{f} \left| \alpha_{r}\right|}
       {\prod_{r=f+1}^{2f} \left| \alpha_{r} \right|}\ .
\end{equation}

On the right hand side of eq.(\ref{eq:VDDF-VLC-RNS}), $X^{\prime-}$
appears only in the form of the vertex operator 
$e^{-ip^{+}X^{\prime-}}$
and $\psi^{\prime-},\tilde{\psi}^{\prime-}$ do not appear. 
Therefore
we can replace $X^{+},\psi^{+},\tilde{\psi}^{+}$ by their expectation
values $-\frac{i}{2}\left(\rho+\bar{\rho}\right),0,0$ in the correlation
function, and vice versa. 
The insertions at $z=z_{I}$ and $\infty$
can be rearranged as 
\begin{eqnarray}
\left|\frac{e^{-\left(1+\alpha\right)
                \left(\sigma^{\prime}-\phi^{\prime}\right)}}
           {\left(\partial^{2}\rho\right)^{1+\frac{\alpha}{4}}}
               \left(z_{I}\right)
\right|^{2} 
& = & \left|\oint_{z_{I}}\frac{dz}{2\pi i}
             \frac{e^{-\sigma^{\prime}}}
                  {\left(\partial\rho\right)^{1+\alpha}}\left(z\right)
            \lim_{w\to z_{I}}
               \left(\left(\partial\rho\right)^{\alpha}
                     e^{\phi^{\prime}}
               \right)  \left(w\right)
           \frac{e^{-\alpha\left(\sigma^{\prime}-\phi^{\prime}\right)}}
                 {\left(\partial^{2}\rho\right)^{\frac{\alpha}{4}}}
                     \left(z_{I}\right)
        \right|^{2}
\nonumber \\
 & \sim & \left|\oint_{z_{I}}\frac{dz}{2\pi i}
                \frac{e^{-\sigma^{\prime}}}
                     {\left(\partial\rho\right)^{1+\alpha}}
                          \left(z\right)
                \lim_{w\to z_{I}}
                   \left(\left(\partial\rho\right)^{\alpha}
                                 e^{\phi^{\prime}}\right)\left(w\right)
            \right|
       \mathcal{O}_{I}\ ,
\nonumber \\
\left|(\partial\rho)^{1+\alpha}e^{\sigma^{\prime}}(\infty)\right|^{2} 
  & = & \left|\left(\sum_{r}\alpha_{r}Z_{r}\right)
               \lim_{z\to\infty}
                  e^{-2\left(\sigma^{\prime}-\phi^{\prime}\right)}
                    \left(z\right)
              \left((\partial\rho)^{\alpha}
                     e^{3\sigma^{\prime}-2\phi^{\prime}}\right)
                 \left(\infty\right)
         \right|^{2}
\nonumber \\
 & \sim & \left|\left(\sum_{r}\alpha_{r}Z_{r}\right)
       \lim_{z\to\infty}e^{-2\left(\sigma^{\prime}-\phi^{\prime}\right)}
                               \left(z\right)
           \right|^{2}
           \mathcal{R}\ .
\label{eq:insertions}
\end{eqnarray}

We also modify the vertex operator
$V_{r}^{\prime\prime\left(-\frac{1}{2},-\frac{1}{2}\right)}$ appearing
on the right hand side of eq.(\ref{eq:VDDF-VLC-RNS}) into
$V_{r}^{\prime\left(-\frac{1}{2},-\frac{1}{2}\right)}$,
which is composed of the free field versions 
$V_{L,r}^{\prime\left(-\frac{1}{2}\right)}$
and $V_{R,r}^{\prime\left(-\frac{1}{2}\right)}$
of the vertex operators
$V_{L,r}^{\left(-\frac{1}{2}\right)}$
and $V_{R,r}^{\left(-\frac{1}{2}\right)}$,
 instead of 
$V_{L,r}^{\prime\prime\left(-\frac{1}{2}\right)}$
and $V_{R,r}^{\prime\prime\left(-\frac{1}{2}\right)}$.
$V_{r}^{\prime\left(-\frac{1}{2},-\frac{1}{2}\right)}$ is defined
to be obtained by applying the picture changing operators to
$V_{r}^{\prime\left(-\frac{3}{2},-\frac{3}{2}\right)}
      (Z_{r},\bar{Z}_{r})$ and
\begin{eqnarray}
V_{r}^{\prime\left(-\frac{1}{2},-\frac{1}{2}\right)}
 & \equiv & X\tilde{X}V^{\prime\left(-\frac{3}{2},-\frac{3}{2}\right)}
\nonumber \\
 & = & 
    \left| -\frac{i}{2}\left(\partial\rho\right)^{\alpha}
            e^{\phi^{\prime}}
            \partial X^{+} \psi^{\prime-}
   \right|^{2}
     V_{r}^{\prime\left(-\frac{3}{2},-\frac{3}{2}\right)}
    + \cdots
\nonumber \\
 & = & 
   \left| -\frac{1}{2} \left(\alpha_{r}\right)^{\alpha}p_{r}^{+}
   \right|^{2}|
   \alpha_{r}|^{-(\alpha+1)}
    V_{L,r}^{\prime\prime\left(-\frac{1}{2}\right)}
    V_{R,r}^{\prime\prime\left(-\frac{1}{2}\right)}+\cdots
\nonumber \\
 & \propto & 
    V_{r}^{\prime\prime\left(-\frac{1}{2},-\frac{1}{2}\right)}
     + \cdots\ .
\end{eqnarray}
Here $\cdots$ denotes the terms which either include derivatives of 
$\partial X^{+}+\frac{i}{2}\partial\rho$,
$\bar{\partial}X^{+}+\frac{i}{2}\bar{\partial}\bar{\rho}$
or are with the fermion numbers 
$\oint\frac{dz}{2\pi i}i\partial H^{\prime}\left(z\right)$,
$\oint\frac{d\bar{z}}{2\pi i}
   i\bar{\partial}\tilde{H}^{\prime}\left(\bar{z}\right)$
bigger than those of 
$V_{r}^{\prime\prime\left(-\frac{1}{2},-\frac{1}{2}\right)}$.
Therefore we can replace 
$V_{r}^{\prime\prime\left(-\frac{1}{2},-\frac{1}{2}\right)}$
in eq.(\ref{eq:VDDF-VLC-RNS}) by 
$V_{r}^{\prime\left(-\frac{1}{2},-\frac{1}{2}\right)}$
without changing the value of the correlation function up to 
a constant multiplicative factor. 

Substituting these into eq.(\ref{eq:VDDF-VLC-RNS}), we get 
\begin{eqnarray}
\lefteqn{
   (2\pi)^{2} \delta\left(\sum_{r=1}^{N}p_{r}^{+}\right)
              \delta\left(\sum_{r=1}^{N}p_{r}^{-}\right)
   e^{-\frac{d-2}{16}\Gamma}
   \prod_{I=1}^{N-2}
     \left| \partial^{2} \rho \left(z_{I}\right) 
     \right|^{-\frac{3}{2}}
   \prod_{r=1}^{N} V_{r}^{\mathrm{LC}}
} \nonumber \\
&  & 
  \sim \left\langle 
          \left| \left(\sum_{r}\alpha_{r}Z_{r}\right)
                 \lim_{z\to\infty}
                 e^{-2\left(\sigma^{\prime}-\phi^{\prime}\right)}
                     \left(z\right)
          \right|^{2}
          \mathcal{R}
       \right.
\nonumber \\
 &  & \hphantom{\quad\sim\quad}
   \times\prod_{I=1}^{N-2}
      \left| \oint_{z_{I}} \frac{dz}{2\pi i}
               \frac{e^{-\sigma^{\prime}}}
                    {\left(\partial\rho\right)^{1+\alpha}}\left(z\right)
               \lim_{w\to z_{I}}
                   \left(\left(\partial\rho\right)^{\alpha} 
                   e^{\phi^{\prime}}\right)\left(w\right)
      \right|   \mathcal{O}_{I}
\nonumber \\
 &  & \hphantom{\quad\sim\quad}
   \left.\times
    \prod_{r=1}^{N}
      \left( |\alpha_{r}|^{-\alpha}
             V_{r}^{\prime\left(p_{L,r},p_{R,r}\right)}
                    \left(Z_{r},\bar{Z}_{r}\right)
      \right)
   \right\rangle _{\mathrm{free}}~,
\label{eq:VDDF-VLC-RNS2}
\end{eqnarray}
where $p_{L,r},p_{R,r}=-\frac{1}{2},-1,-\frac{3}{2}$ indicate the
picture of the vertex operator. The choice of picture is obvious from
eq.(\ref{eq:VDDF-VLC-RNS}). To this equation we can easily apply
the formula~(\ref{eq:supergeneral}) and express the right hand side
using the $X^{\pm}$ CFT and the unprimed ghost fields. Substituting
it into eq.(\ref{eq:FN}), we obtain 
\begin{equation}
F_{N} 
 \sim  
  \left\langle 
      \left|\partial\rho c\left(\infty\right)\right|^{2}
      \prod_{I}\left|\oint_{z_{I}}\frac{dz}{2\pi i}
                        \frac{b}{\partial\rho}\left(z\right)
                     e^{\phi}T_{F}^{\mathrm{LC}}\left(z_{I}\right)
                \right|^{2}
      \prod_{r=1}^{N}\mathcal{S}_{r}^{-1}
      \prod_{r=1}^{N}V_{r}^{\left(p_{L,r},p_{R,r}\right)}
          \left(Z_{r},\bar{Z}_{r}\right)\right\rangle.
\label{eq:FNunprime}
\end{equation}
Here $\left\langle \cdots\right\rangle $ denotes the correlation
function of the CFT for the longitudinal 
and transverse variables and the super-reparametrization ghosts. 
$V_{r}^{\left(p_{L,r},p_{R,r}\right)}$ is
the unprimed field version of 
$V_{r}^{\prime\left(p_{L,r},p_{R,r}\right)}$
and it is BRST invariant. 
$\mathcal{S}_{r}^{-1}$ 
is defined as 
\begin{equation}
\mathcal{S}_{r}^{-1}
  \equiv
  \oint_{z_{I^{\left(r\right)}}} \frac{d\mathbf{z}}{2\pi i}
         D\Phi\left(\mathbf{z}\right)
  \oint_{\bar{z}_{I^{\left(r\right)}}} \frac{d\bar{\mathbf{z}}}{2\pi i}
         \bar{D}\Phi\left(\bar{\mathbf{z}}\right)
  e^{\frac{d-10}{16}\frac{i}{p_{r}^{+}}\mathcal{X}^{+}}
         \left(\mathbf{z},\bar{\mathbf{z}}\right)\ ,
\end{equation}
which can be shown to be the inverse of $\mathcal{S}_{r}$ 
in eq.(\ref{eq:mathcalSr})
by replacing $\mathcal{X}^{+}$ by its expectation value. 
$\mathcal{S}_{r}^{-1}$
coincides with the BRST invariant form of 
$e^{\frac{d-10}{16}\frac{i}{p_{r}^{+}}X^{+}}
    (z_{I^{(r)}},\bar{z}_{I^{(r)}})$
introduced in Ref.~\cite{Baba:2009zm}.

\subsection{BRST invariant form of the amplitudes }

\label{sec:picturechanging}

In eq.(\ref{eq:FNunprime}), the right hand side is expressed by the
variables in the conformal gauge, but it is not manifestly BRST invariant.
In order to get a BRST invariant form of the amplitudes, we would
like to show that  $e^{\phi}T_{F}\left(z_{I}\right)$ 
in eq.(\ref{eq:FNunprime})
can be turned into the picture changing operator $X(z_{I})$ and 
\begin{equation}
F_{N}  \sim 
   \left\langle \left|\partial\rho c\left(\infty\right)\right|^{2}
       \prod_{I}\left|\oint_{z_{I}}\frac{dz}{2\pi i}
                          \frac{b}{\partial\rho}\left(z\right)
                       X\left(z_{I}\right)
                 \right|^{2}
        \prod_{r=1}^{N}\mathcal{S}_{r}^{-1}
        \prod_{r=1}^{N}V_{r}^{\left(p_{L,r},p_{R,r}\right)}
   \right\rangle ~.
\label{eq:FNpic}
\end{equation}

\subsubsection*{picture changing operator}

Let us introduce a nilpotent fermionic charge $Q$ defined as 
\begin{equation}
Q \equiv
  \oint\frac{dz}{2\pi i}
    \left[ -\frac{1}{4} \frac{b}{\partial\rho}
            \left(iX_{L}^{+} - \frac{1}{2} \rho \right)
           +\frac{1}{2}
            \frac{e^{-\phi}\partial\xi}{\partial\rho}\psi^{+}
     \right]  \left(z\right)\ .
\end{equation}
Here we define $X_{L}^{+}\left(z\right)$ so that 
\begin{equation}
\left(iX_{L}^{+}-\frac{1}{2}\rho\right)  \left(z\right)
  =\int_{\infty}^{z} dz^{\prime}
     \left(i\partial X^{+}-\frac{1}{2}\partial\rho\right)
          \left(z^{\prime}\right)\ .
\end{equation}
One can show 
\begin{eqnarray}
\oint_{z_{I}}\frac{dz}{2\pi i}
  \frac{b}{\partial\rho}\left(z\right) X(z_{I}) 
& = & -\oint_{z_{I}} \frac{dz}{2\pi i}
   \frac{b}{\partial\rho}\left(z\right)
   e^{\phi}T_{F}^{\mathrm{LC}}\left(z_{I}\right)
\nonumber \\
 &  & {}+\left[Q\,,\,
               \oint_{z_{I},w} \frac{dz}{2\pi i}
                 \frac{b}{\partial\rho} \left(z\right)
               \oint_{z_{I}}\frac{dw}{2\pi i}
                  \frac{A(w)}{w-z_{I}}e^{\phi} \left(z_{I}\right)
          \right]
\nonumber \\
 &  & {}+\frac{1}{4}
         \oint_{z_{I},w}\frac{dz}{2\pi i}
           \frac{b}{\partial\rho}\left(z\right)
         \oint_{z_{I}}\frac{dw}{2\pi i}
           \frac{\partial\rho\psi^{-}\left(w\right)}{w-z_{I}}
           e^{\phi}\left(z_{I}\right),
\label{eq:XQexact}
\end{eqnarray}
where 
\begin{eqnarray}
A\left(w\right) 
 & \equiv & 
  {}-i\partial X^{+}\partial\rho\eta e^{\phi}\left(w\right)
     -2\partial\left(\partial\rho c\right)\psi^{-}\left(w\right)
\nonumber \\
&  & 
  {}-\frac{d-10}{4}
    i \left[ 
         \left( \frac{5\left(\partial^{2}X^{+}\right)^{2}}
                     {4\left(\partial X^{+}\right)^{3}}
                -\frac{\partial^{3}X^{+}}
                      {2\left(\partial X^{+}\right)^{2}}
         \right)
           \left(-2\partial\rho\eta e^{\phi}\right)
        -\frac{2\partial^{2}X^{+}}{\left(\partial X^{+2}\right)}
          \partial\left(-2\partial\rho\eta e^{\phi}\right)
\right. \nonumber \\
&  & \hphantom{\quad-\frac{d-10}{4}i}
   \left.
   {}+ \frac{\partial^{2}\left(-2\partial\rho\eta e^{\phi}\right)}
            {\partial X^{+}}
     -\frac{ \left(-2\partial\rho\eta e^{\phi}\right)
             \partial\psi^{+}\partial^{2}\psi^{+}}
           {2\left(\partial X^{+}\right)^{3}}
\right]   \left(w\right)\ .
\end{eqnarray}
Substituting eq.(\ref{eq:XQexact}) into the right hand side 
of eq.(\ref{eq:FNpic})
and comparing it with that of eq.(\ref{eq:FNunprime}), 
one can see that in order to prove eq.(\ref{eq:FNpic}),
one should show that the second and the third terms
on the right hand side of eq.(\ref{eq:XQexact}) do not contribute
to the correlation function.

One can prove the third term does not contribute to the correlation 
function by describing the insertion~(\ref{eq:XQexact})
in terms of the free variables. 
The proof is given in appendix \ref{sec:method2}.\footnote{
Actually this term has the same structure as the third term on the right
hand side of eq.(3.25) in Ref.~\cite{Baba:2009zm}. 
In Ref.~\cite{Baba:2009zm},
we have given another proof that the contributions of such terms vanish.}
The second term is $Q$-exact. We can therefore prove that this is
also irrelevant, by showing that $Q$ (anti)commutes with all the
operators in the correlation function (\ref{eq:FNpic}). 
It is straightforward to show that $Q$ (anti)commutes with 
the vertex operators. Moreover,
$Q$ (anti)commutes with other insertions: 
\begin{eqnarray}
 &  & \left\{ Q\,,\,\partial\rho c\left(\infty\right)\right\} 
      =-\frac{1}{4} \left(iX_{L}^{+}-\frac{1}{2}\rho\right)
                           \left(\infty\right)
      =0~,
      \qquad
      \left\{ Q\,,\,
              \oint_{z_{I}}\frac{dz}{2\pi i}\frac{b}{\partial\rho}
      \right\} =0~,
\nonumber \\
 &  & \left[Q\,,\, e^{\phi}T_{F}^{\mathrm{LC}}\left(z_{I}\right)
      \right]
      =\left[Q\,,\,\oint_{z_{I}}\frac{dw}{2\pi i}\frac{1}{w-z_{I}}
                    \partial\rho\psi^{-}\left(w\right)
                    e^{\phi}\left(z_{I}\right)
        \right]=0\ .
\end{eqnarray}
Thus we obtain the expression (\ref{eq:FNpic}) for $F_{N}$.

\subsubsection*{BRST invariant form}

By deforming the contour of 
$\oint_{z_{I}}\frac{dz}{2\pi i}\frac{b}{\partial\rho}\left(z\right)$
in eq.(\ref{eq:FNpic})
as was done in Ref.~\cite{Baba:2009zm}, we can obtain a manifestly
BRST invariant form of the amplitude $\mathcal{A}_{N}$:
\begin{eqnarray}
&&
\mathcal{A}_{N} 
 \sim 
  \int \prod_{\mathcal{I}=1}^{N-3} d^{2}\mathcal{T}_{\mathcal{I}}
  \left\langle 
     \prod_{\mathcal{I}=1}^{N-3}
        \left[ \oint_{C_{\mathcal{I}}}\frac{dz}{2\pi i}
                  \frac{b}{\partial\rho}(z)
               \oint_{C_{\mathcal{I}}}\frac{d\bar{z}}{2\pi i}
                   \frac{\tilde{b}}{\bar{\partial}\bar{\rho}}(\bar{z})
          \right]
     \prod_{I} \left| X \left(z_{I}\right) \right|^{2}
\right. \nonumber \\
&& \hspace{9em} \left. \times
     \prod_{r=1}^{N} \mathcal{S}_{r}^{-1}
     \prod_{r=1}^{N}V_{r}^{\left(p_{L,r},p_{R,r}\right)}
  \right\rangle \ .
\end{eqnarray}
Here $C_{\mathcal{I}}$ denotes a contour which goes around the 
$\mathcal{I}$th internal propagator. 
 Compared with the form of the tree amplitudes in the critical case, 
the difference is in the insertions of $\mathcal{S}_{r}^{-1}$. These
insertions are peculiar to the noncritical strings~\cite{Baba:2009ns}.

\section{Conclusions and discussions \label{sec:Conclusions-and-discussions}}

In this paper, we have formulated a free field description of the
$X^{\pm}$ CFT combined with the reparametrization ghosts, and provided
a formula to express the correlation functions in terms of the free
variables. Since the $X^{\pm}$ CFT is an interacting theory, it is
not straightforward to construct spin fields and thus the vertex operators
in the Ramond sector. We have given the spin fields 
via the free variables, and thereby we have
constructed the BRST invariant vertex operators in the Ramond sector.
We have shown how one can calculate tree amplitudes with the external
lines in the (R,R) sector as well as those in the (NS,NS) sector in
the noncritical string theory using these vertex operators in the
conformal gauge. 

We study such noncritical string field theories, in order to dimensionally
regularize the string field theory to deal with the divergences of
the theory~\cite{Baba:2009kr,Baba:2009zm}. One occasion in which
such regularization is useful is when we deal with the contact term
problem~\cite{Greensite:1986gv,Greensite:1987sm,%
              Greensite:1987hm,Green:1987qu,Wendt:1987zh}.
In the light-cone gauge superstring field theory, even tree amplitudes
are divergent because of the existence of the supercurrent insertions
at the interaction points. Using the results obtained in this paper,
we can show that the dimensional regularization can be employed to
deal with the contact term problem for the tree amplitudes when the
external lines are in the (R,R) and the (NS,NS) sectors.

In order to generalize our regularization scheme to the amplitudes
involving external lines in the (R,NS) and the (NS,R) sectors, there
are several issues to be resolved. As we have pointed out, if we take
$d\ne10$ naively, we get a theory with no spacetime fermions. 
In order to
deal with this problem, we need to modify the worldsheet theory. 
Moreover
the dimensional regularization in usual field theory
for point particles has some problems
in treating fermions. We encounter similar problems when we try to
apply the regularization to superstring field theory. 
We will discuss these points elsewhere.

Another thing to be examined is the Green-Schwarz formalism. As was
commented in Ref.~\cite{Kazama:2010ys}, the results 
in Ref.~\cite{Baba:2009ns} seems to be useful in constructing 
vertex operators in the semi-light-cone
gauge formulation of the Green-Schwarz formalism, recently re-examined
in Refs.~\cite{Kazama:2010ys,Berkovits:2004tw,Aisaka:2005vn,
               Kunitomo:2006gr,Kazama:2008as}.
Moreover, the similarity transformation given 
in Ref.~\cite{Kazama:2010ys} looks similar to 
the field redefinition~(\ref{eq:free})~\cite{Kazama2010}.
It will be interesting to examine how the results in this paper are
related to these developments.
\section*{Acknowledgements}

We would like to thank Y.~Kazama and F.~Sugino
for valuable discussions and comments.
We are also grateful to the Yukawa Institute for Theoretical Physics 
at Kyoto University, where part of this work was done during 
the YITP-W-10-13 on ``String Field Theory and Related Aspects",
which was supported  by the Grant-in-Aid for the Global COE Program
``The Next Generation of Physics, Spun from Universality and
Emergence" from the Ministry of Education, Culture, Sports, Science and
Technology (MEXT) of Japan.
This work was supported in part by 
Grant-in-Aid for Scientific Research~(C) (20540247) from MEXT.


\appendix

\section{String field theory action in $d$ dimensions}
\label{sec:SFTaction}

In this appendix, we explain some details of the action of the light-cone
gauge superstring field theory in noncritical dimensions.

We represent the string field $|\Phi_{\lambda}(t)\rangle$ by a wave
function for the bosonic zero modes $(t,\alpha,\vec{p})$ and a Fock
state for the other modes. We denote the integration measure of the
momentum zero modes of the $r$th string by $dr$, which is defined
as 
\begin{equation}
dr=\frac{\alpha_{r}d\alpha_{r}}{4\pi}
    \frac{d^{d-2}p_{r}}{(2\pi)^{d-2}}~.
\end{equation}
The string fields are taken to be GSO even and satisfy 
the level-matching condition. 
The subscript $\lambda$ of the string field labels the
sector to which the string field belongs. 
As was stated in section~\ref{sec:Amplitudes},
we concentrate on the strings in the (NS,NS) and the (R,R) sectors.
Therefore the subscript $\lambda$ takes only (NS,NS) and (R,R) and
the string fields $\Phi_{\lambda}$ are Grassmann
even. The action for the string fields in these sectors of the light-cone
gauge superstring field theory in $d$ dimensions $(d\neq10)$ takes
the form 
\begin{eqnarray}
S & = & \int dt
  \left[\frac{1}{2} \sum_{\lambda}
        \int d1d2
          \left\langle R_{\lambda}\left(1,2\right)|
              \Phi_{\lambda}(t)\right\rangle _{1}
         \left(i\frac{\partial}{\partial t}
                -\frac{L_{0}^{\mathrm{LC}(2)}
                        +\tilde{L}_{0}^{\mathrm{LC}(2)}-\frac{d-2}{8}}
                       {\alpha_{2}}
          \right)
          \left|\Phi_{\lambda}(t)\right\rangle _{2}
   \right.
\nonumber \\
 &  & \hphantom{\int dt\frac{1}{2}}
   {}+\frac{2g}{3}\int d1d2d3
       \left\langle V_{3}\left(1_{\mathrm{NSNS}},2_{\mathrm{NSNS}},
                               3_{\mathrm{NSNS}}\right)
       |\Phi_{\mathrm{NSNS}}(t)\right\rangle _{1}
       \left|\Phi_{\mathrm{NSNS}}(t)\right\rangle _{2}
       \left|\Phi_{\mathrm{NSNS}}(t)\right\rangle _{3}
\nonumber \\
 &  & \hphantom{\int dt\frac{1}{2}}
    \left.
       \vphantom{\left(i 
                 \frac{\partial}{\partial t}
                       -\frac{L_{0}^{\mathrm{LC}\left(2\right)}
                       +\tilde{L}_{0}^{\mathrm{LC}\left(2\right)}
                       -\frac{d-2}{8}}
                      {\alpha_{2}}\right)}
      {}+2g\int d1d2d3
         \left\langle V_{3}\left(1_{\mathrm{NSNS}},2_{\mathrm{RR}},
                                 3_{\mathrm{RR}}\right)
         |\Phi_{\mathrm{NSNS}}(t)\right\rangle _{1}
         \left|\Phi_{\mathrm{RR}}(t)\right\rangle _{2}
         \left|\Phi_{\mathrm{RR}}(t)\right\rangle _{3}
    \right]\ .
\label{eq:sftaction}
\end{eqnarray}
Here $\langle R_{\lambda}(1,2)|$ denotes the reflector for the string
fields in sector $\lambda$. 
$\langle V_{3}(1_{\lambda_{1}},2_{\lambda_{2}},3_{\lambda_{3}})|$
denotes the interaction vertex for the three strings 
in sector $\lambda_{r}$ $(r=1,2,3)$. 
This is invariant under the permutation of the string
fields and takes the form 
\begin{eqnarray}
\langle V_{3}(1_{\lambda_{1}},2_{\lambda_{2}},3_{\lambda_{3}})| 
  & = & 4\pi\delta\left(\sum_{r=1}^{3}\alpha_{r}\right)
        (2\pi)^{d-2}\delta^{d-2}\left(\sum_{r=1}^{3}p_{r}\right)
\nonumber \\
 &  & \times
   \langle V_{3}^{\mathrm{LPP}}(1_{\lambda_{1}},2_{\lambda_{2}},
                                 3_{\lambda_{3}})|
   P_{123}e^{-\Gamma^{[3]}(1,2,3)}~.
\end{eqnarray}
Here $\langle V_{3}^{\mathrm{LPP}}(1,2,3)|$ denotes 
the LPP vertex~\cite{LeClair:1988sp},
which satisfies eq.(A.5) of Ref.~\cite{Baba:2009zm}. 
$\Gamma^{[3]}(1,2,3)$ and $P_{123}$ are defined 
in eqs.(A.4) and (A.6) of Ref.~\cite{Baba:2009zm} respectively.


\section{Correlation functions of $\psi^{-}$}

\label{sec:method2}

In this appendix, we show that the third term on the right hand side
of eq.(\ref{eq:XQexact}) does not contribute the correlation function.
In terms of free fields, the third term on the right hand side of 
eq.(\ref{eq:XQexact}) turns out to be 
\begin{equation}
\frac{1}{\partial^{2}\rho(z_{I})}
        \frac{1}{4}
          \oint_{z_{I}}\frac{dw}{2\pi i}
            \frac{\partial\rho\psi^{-}\left(w\right)}
                 {w-z_{I}}
e^{-\sigma^{\prime}+\phi^{\prime}}\left(z_{I}\right)
\mathcal{O}_{I}\ ,
\label{eq:XQexactfree}
\end{equation}
where $\psi^{-}\left(w\right)$ is written
using the free fields as 
\begin{equation}
\psi^{-}\left(w\right)
 =\psi^{\prime-}\left(w\right)+\delta\psi^{-}\left(w\right)\ .
\label{eq:psi-free}
\end{equation}
The explicit form of $\delta\psi^{-}\left(w\right)$ can be deduced
from eq.(\ref{eq:Xprime-}) but we do not need it here. 

We would like to show that the integral 
$\oint_{z_{I}}\frac{dw}{2\pi i}
   \frac{\partial\rho\psi^{-}\left(w\right)}{w-z_{I}}$
does not contribute to the correlation function. 
It can have nonzero contributions 
when $\psi^{-}\left(w\right)$ has singularities at $w=z_{I}$. 
Using the expression (\ref{eq:psi-free}), one can see
that such singularities come either from contracting $\psi^{\prime-}$
with $\psi^{+}$ included in $\mathcal{O}_{I}$ 
or from $\delta\psi^{-}\left(w\right)$.
$\delta\psi^{-}\left(w\right)$ involves factors 
$\left(\partial X^{+}\right)^{-n}\left(w\right)$,
which has the expectation value 
$\left(-\frac{i}{2}\partial\rho\right)^{-n}\left(w\right)$
and singular at $w=z_{I}$. 
In order to give a nonvanishing contribution,
$\psi^{\prime-}$ contained in 
$\oint_{z_{I}}\frac{dw}{2\pi i}
  \frac{\partial\rho\psi^{-}\left(w\right)}{w-z_{I}}$
should be contracted with $\psi^{+}$ contained in $\mathcal{O}_{I}$
and not in $\mathcal{O}_{J}$ with $J\neq I$. 
Since the Grassmann odd quantities in $\mathcal{O}_{I}$ 
and $\delta\psi^{-}$ are made from $\psi^{+}$ and $\beta c$, 
these terms necessarily involves
derivatives of $\psi^{+}$ and $\beta c$. 
If we take contractions of all $\psi^{\prime-}$ 
with appropriate $\psi^{+}$'s, the resulting
contributions of the second term on the right hand side of 
eq.(\ref{eq:XQexactfree})
to the correlation functions can be seen to vanish because of the
conservation of the fermion number 
$\oint\frac{dz}{2\pi i}i\partial H^{\prime}$
and the $bc$ ghost number.

\bibliographystyle{utphys}
\bibliography{private,SFTSep20_10}

\end{document}